\pdfoutput=1
\newcommand*{\ATLASLATEXPATH}{}
\documentclass[cernpreprint, texlive=2016, UKenglish]{atlasdoc}

\DeclareOldFontCommand{\rm}{\normalfont\rmfamily}{\mathrm}
\DeclareOldFontCommand{\bf}{\normalfont\bfseries}{\mathbf}
\DeclareOldFontCommand{\sc}{\normalfont\scshape}{\@nomath\sc}

\usepackage{\ATLASLATEXPATH atlaspackage}
\usepackage[italic]{hepnames}
\usepackage{\ATLASLATEXPATH atlasbiblatex}
\usepackage{\ATLASLATEXPATH atlasphysics}
\usepackage{multirow}
\usepackage{lscape}
\usepackage{rotating}
\usepackage{textcomp}
\usepackage{todonotes}
\usepackage{onelepton2016_Moriond-defs}

\graphicspath{{logos/}{figures/}}

\AtlasRefCode{SUSY-2016-12}

\PreprintIdNumber{CERN-EP-2017-140}

\AtlasJournal{PRD}

\AtlasJournalRef{Phys. Rev. D 96 (2017) 112010}
\AtlasDOI{10.1103/PhysRevD.96.112010}

\newcommand{\intlumi}{36.1\,\ifb}
\AtlasAbstract{
The results of a search for squarks and gluinos in final states with an
isolated electron or muon, multiple jets and large missing transverse momentum
using proton--proton collision data at a center-of-mass energy of
$\sqrt{s}$ = 13~\TeV~are presented.
The dataset used was recorded during 2015 and 2016 by the ATLAS experiment at the
Large Hadron Collider and corresponds to an integrated luminosity of
\intlumi. No significant excess beyond the expected background is
found.
Exclusion limits at 95\% confidence level are set in a number of supersymmetric
scenarios, reaching masses up to 2.1~\TeV~for gluino pair production and up to
1.25~\TeV~for squark pair production.
}

\AtlasTitle{Search for squarks and gluinos in events with an isolated lepton, jets, and
missing transverse momentum at $\sqrt{s}$ = 13~\TeV~with the ATLAS detector}
\author{The ATLAS Collaboration}
\hypersetup{pdftitle={ATLAS draft},pdfauthor={The ATLAS Collaboration}}

\begin{document}
\maketitle

\section{Introduction}
\label{sec:intro}
Supersymmetry (SUSY)
\cite{Golfand:1971iw,Volkov:1973ix,Wess:1974tw,Wess:1974jb,Ferrara:1974pu,Salam:1974ig}
is a theoretical framework of physics beyond the Standard Model (SM) which
predicts for each SM particle the existence of a supersymmetric partner
(sparticle) differing by half a unit of spin. The partner particles of the SM
fermions (quarks and leptons) are the scalar squarks ($\tilde{q}$) and
sleptons~($\tilde{\ell}$).
In the boson sector, the supersymmetric partner of the gluon is the fermionic
gluino ($\tilde{g}$), whereas the supersymmetric partners of the Higgs
(higgsinos) and the electroweak gauge bosons (winos and bino) mix to form
charged mass eigenstates (charginos) and neutral mass eigenstates (neutralinos).
In the minimal supersymmetric extension of the Standard Model
(MSSM)~\cite{Fayet:1976et,Fayet:1977yc} two scalar Higgs doublets along with
their higgsino partners are necessary, resulting in four chargino states
($\tilde{\chi}^{\pm}_{1,2}$) and four neutralinos
($\tilde{\chi}^{0}_{1,2,3,4}$).
SUSY addresses the SM hierarchy
problem~\cite{Sakai:1981gr,Dimopoulos:1981yj,Ibanez:1981yh,Dimopoulos:1981zb}
provided that the masses of at least some of the supersymmetric particles (most
notably the higgsinos, the top squarks and the gluinos) are near the \TeV\
scale.

In R-parity-conserving SUSY~\cite{Farrar:1978xj}, gluinos or squarks are pair
produced at the Large Hadron Collider (LHC) via the strong interaction and decay either directly or
via intermediate states to the lightest supersymmetric particle (LSP).
The LSP, which is assumed to be the lightest neutralino~(\ninoone) in this paper,
is stable and weakly interacting, making it a candidate for
dark matter~\cite{Goldberg:1983nd,Ellis:1983ew}. 

\begin{figure}[b]
\centering
\includegraphics[width=0.3\textwidth]{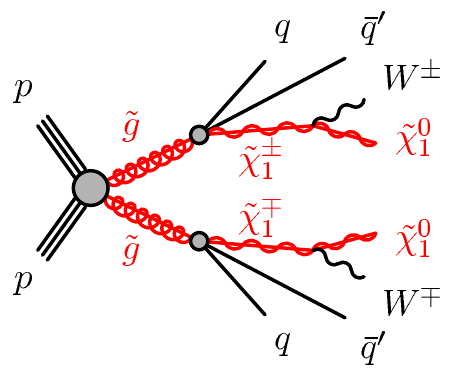}
\includegraphics[width=0.3\textwidth]{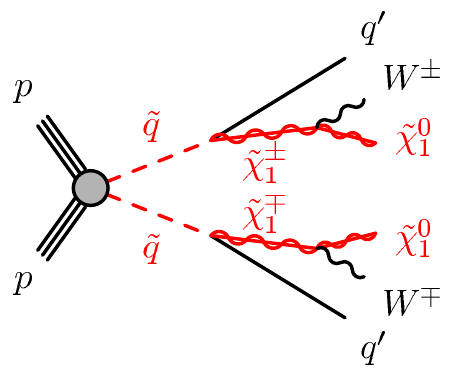}
\includegraphics[width=0.3\textwidth]{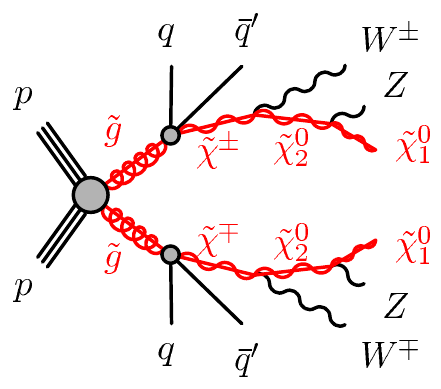}
\caption{The decay topologies of the simplified signal models considered in this
search:
gluino one-step (left), squark one-step (middle), and gluino two-step (right).}
\label{fig:feynman}
\end{figure}

The decay topologies targeted in this paper are largely inspired by decay chains that could
be realized in the pMSSM scenario, which is a two-dimensional subspace of the 
19-parameter phenomenological Minimal Supersymmetric Standard Model (pMSSM)~\cite{Djouadi:1998di,Berger:2008cq}. Four SUSY models with gluino or squark pair production
and different decay topologies are considered. 
The first two models, referred to as the gluino and squark one-step models for
the rest of this paper, are SUSY simplified
models~\cite{Alwall:2008ve,Alwall:2008ag,Alves:2011wf} in which pair-produced
gluinos or squarks decay via the lightest chargino~(\chinoonepm) to the LSP\@.  
In the model with gluino production, the gluino decays to the lightest
chargino and two SM quarks via $\gluino \rightarrow q\bar{q}' \chinoonepm$, 
as illustrated in Figure~\ref{fig:feynman} (left). 
The gluino decay is assumed to proceed via virtual first- and second-generation
squarks, hence no bottom or top quarks are produced in the
simplified model. The chargino then decays to the LSP by emitting an on- or
off-shell $W$ boson, $\chinoonepm \rightarrow W^{(*)\pm}\ninoone$, depending on the available phase space. 
In the MSSM this decay chain is realized when the gluino decays, via a
virtual squark that is the partner particle of the left-handed SM quark, to the
chargino with a dominant wino component. 
In the squark production model, the
squark decays to the chargino via $\squark \rightarrow q' \chinoonepm$, followed
by the same chargino decay,
as illustrated in Figure~\ref{fig:feynman} (middle). 

The third model, referred to as the gluino two-step model for the rest of this
paper, assumes gluino pair production with a
subsequent decay to the chargino via $\gluino \rightarrow q \bar{q}'
\chinopm$. The chargino then decays via emission of an on- or off-shell $W$
boson to the second lightest neutralino according to $\chinopm \rightarrow
W^{\pm} \ninotwo$. In the last step of the cascade, the second lightest
neutralino decays via emission of a $Z$ boson to the LSP. The decay chain
of this signal model is illustrated in Figure~\ref{fig:feynman}~(right). The model is used as a proxy for SUSY scenarios with many decay products
in the final state. Within the MSSM, additional decay modes lead to a
significant reduction in the cross-section times branching fraction for this
particular decay.

Finally, the fourth set of SUSY models, 
the pMSSM model, is selected to have a bino-dominated neutralino as the LSP,
kinematically accessible gluinos, and a higgsino-dominated multiplet at intermediate mass.
The higgsino multiplet contains two neutralinos (the \ninotwo{} and
\ninothree{}) and a chargino. The decays proceed predominantly via virtual
third-generation supersymmetric quarks due to their enhanced couplings with the
higgsinos. Examples of dominant characteristic decay chains of this
model for $\mchino \lsim 500\GeV$ and $\mgluino \gsim 1200\GeV$ are
$\gluino  \to t \bar{t} \neuttwothree$ and $\gluino  \to t \bar{b} \chargino$,
 with \neuttwothree decaying to $Z / h$  $\neut$ and \chargino to
$W^{\pm} \neut$. 

In this search, the experimental signature consists of a lepton
(electron or muon), several jets, and missing transverse momentum (\met) from
the undetectable neutralinos and neutrino(s). Depending on the sparticle masses
of the model considered, different amounts of energy are available in their
decays. Therefore, the number of leptons and jets in the final state, as well as
their kinematic properties, depend on the mass spectrum in the model of
interest. Four signal regions with jet multiplicities ranging from two to six are defined to provide sensitivity to
a broad range of mass spectra in the gluino and squark one-step models. For the
two-step and pMSSM models, a dedicated signal region requiring nine jets is
constructed to take advantage of the large jet multiplicities in these models. 
In each signal region, the event yield is compared with the SM prediction, which is
estimated using a combination of simulation and observed data in control regions.

The search presented in this paper uses the ATLAS data collected in
proton--proton collisions at the LHC during 2015 and 2016
at a center-of-mass energy of 13\,\TeV,
corresponding to an integrated luminosity of \intlumi.
The analysis extends previous ATLAS searches with similar event selections
performed with data collected in 2010--2012 (LHC Run-1)~\cite{Aad:2015mia}
and in 2015 (Run-2)~\cite{Aad:2016qqk}, at center-of-mass energies of
8\,\TeV and 13\,\TeV, respectively. 
Similar searches for gluinos and squarks with decays via intermediate
supersymmetric particles were performed by the CMS Collaboration in Run
1~\cite{Chatrchyan:2013fea,Chatrchyan:2014lfa} and Run
2~\cite{Khachatryan:2016xdt,Sirunyan:2017uyt,Sirunyan:2017cwe}. The results of all Run-1 ATLAS 
searches targeting squark and gluino pair production are summarized in Ref.~\cite{Aad:2015iea}. 
The same SUSY models considered in this paper were also targeted in other Run-2 ATLAS 
searches using different experimental signatures~\cite{Aaboud:2016zdn,Aad:2016jxj,Aaboud:2017dmy}.

This paper is structured as follows. 
After a brief description of the ATLAS detector in Section~\ref{sec:detector}, the simulated data samples for the background and signal processes used in the analysis as well as the dataset and the trigger strategy are detailed in Section~\ref{sec:samples}. 
The reconstructed objects and quantities used in the analysis are described in Section~\ref{sec:objects} and the event selection is presented in Section~\ref{sec:selection}. 
The background estimation and the systematic uncertainties associated with the expected event yields are discussed in Sections~\ref{sec:background} and~\ref{sec:systematics}, respectively.
Finally, the results of the analysis are presented in Section~\ref{sec:result},
and are followed by a conclusion.

\section{ATLAS detector}
\label{sec:detector}
\interfootnotelinepenalty=10000

ATLAS~\cite{PERF-2007-01} is a general-purpose detector with a forward-backward
symmetric design that provides almost full solid angle coverage around the
interaction point.\footnote{%
    ATLAS uses a right-handed coordinate system with its
    origin at the nominal interaction point (IP) in the center of the detector and
    the $z$-axis along the beam pipe. The $x$-axis points from the IP to the center
    of the LHC ring, and the $y$-axis points upward. Cylindrical coordinates
    $(r,\phi)$ are used in the transverse plane, $\phi$ being the azimuthal angle
    around the $z$-axis. The pseudorapidity is defined in terms of the polar angle
    $\theta$ as $\eta=-\ln\tan(\theta/2)$. Rapidity is defined as $y = 0.5 \ln [(E+p_{z})/(E-p_{z})]$
    where $E$ denotes the energy and $p_{z}$ is the component of the momentum along
    the beam direction.}
The main components are the inner detector (ID), which is surrounded by
a superconducting solenoid providing a $2$ T axial magnetic field, the calorimeter
system, and the muon spectrometer (MS), which is immersed in a magnetic field
generated by three large superconducting toroidal magnets. The ID provides track
reconstruction within $|\eta| < 2.5$, employing pixel detectors close to the beam pipe, silicon microstrip detectors at intermediate radii, and a straw-tube tracker with particle identification capabilities based on transition radiation at radii up to $1080$ mm. The innermost pixel detector layer, the insertable B-layer~\cite{B-layerRef}, was added during the shutdown between LHC Run 1 and Run 2, at a radius of $33$ mm around a new, narrower, beam pipe. The calorimeters cover $|\eta| < 4.9$. The forward region ($3.2 < |\eta| < 4.9$) is instrumented with a liquid-argon (LAr) calorimeter for both the
electromagnetic and hadronic measurements. In the central region, a
lead/LAr electromagnetic calorimeter covers $|\eta| < 3.2$,
while the hadronic calorimeter uses two different detector technologies, with
scintillator tiles ($|\eta| < 1.7$) or liquid argon ($1.5 < |\eta| < 3.2$) as
the active medium. The MS consists of three layers of precision tracking chambers providing coverage
over $|\eta| < 2.7$, while dedicated fast chambers allow triggering over $|\eta|
< 2.4$. The ATLAS trigger system used for real-time event
selection~\cite{atlastrigger2015} consists of a hardware-based first-level trigger and a software-based high-level trigger.

\section{Simulated event samples and data samples}
\label{sec:samples}

Three simplified SUSY signal models and
a set of pMSSM scenarios are considered in this search.
Gluinos or squarks are assumed to be
produced in pairs ($\gluino\gluino$ or $\squark\bar\squark$). In the
case of the simplified models, 100\% branching ratios to the decay of interest
are assumed.

The gluino/squark one-step simplified models have three free parameters: 
the masses of the gluino or squark (\mglsq), the lightest chargino (\mchino), 
and the lightest neutralino (\mnino).
Other sparticles that do not appear in the decay chain 
are set to have a high mass. To probe a broad
range of SUSY mass spectra, two model parameterizations are
considered.
In the first type,
\mglsq\ and the mass ratio
$x \equiv (\mchino - \mnino)/(\mglsq - \mnino)$
are free parameters, while \mnino\ is fixed to $60\GeV$.
In the second type, \mglsq\ and \mnino\ are free parameters,
while \mchino\ is fixed by setting $x = 1/2$.
For the rest of this paper, the former type is referred to as variable-$x$
and the latter one is referred to as $x = 1/2$.

The gluino two-step simplified model has two free parameters that are varied
to probe different mass configurations: the masses of the gluino (\mgluino) and
the lightest neutralino (\mnino).
The masses of the lightest chargino and the second-lightest neutralino are constrained to be $\mchino=(\mgluino + \mnino)/2$ and $\mninotwo=(\mchino + \mnino)/2$, respectively.
All other sparticles are kinematically inaccessible. 

In the pMSSM scenario, the sparticle masses are varied by scanning the gluino
mass parameter $M_3$ (related to \mgluino) and the bilinear Higgs mass parameter
$\mu$ (related to \mchino\ and \mninotwo). The scan ranges are $690 \GeV < M_3 <
2140 \GeV$ and $-770 \GeV < \mu < -160 \GeV$. The bino mass
parameter $M_1$ (related to \mnino) was set to $60\GeV$.
The remaining model parameters, defined in Ref.~\cite{Aad:2015baa}, are set to
$M_A=M_2=3\TeV,$ $A_\tau=0$, $\tan\beta=10$, and
$A_t=A_b=m_{\tilde{{\rm
L}_{\rm
L}}_{(1,2,3)}}=m_{(\tilde{e_{\rm
R}},\tilde{\mu_{\rm
R}},\tilde{\tau_{\rm
R}})}=m_{\tilde{Q_{\rm
L}}_{(1,2,3)}}=m_{(\tilde{u_{\rm
R}},\tilde{c_{\rm
R}},\tilde{t_{\rm
R}})}=m_{(\tilde{d_{\rm R}},\tilde{s_{\rm R}},\tilde{b_{\rm R}})}=5\TeV$, such
that the mass of the lightest Higgs boson is compatible with 125~\GeV~and all
other sparticles are kinematically inaccessible. Mass spectra consistent with
electroweak symmetry breaking were generated using
{\sc SOFTSUSY}~3.4.0~\cite{Allanach:2001kg} and the decay branching ratios were
calculated with {\sc SDECAY/HDECAY}~1.3b/3.4~\cite{Djouadi:2006bz}.

The signal samples were generated at leading order (LO)
using \MADGRAPH 2.2.2~\cite{Alwall:2014hca}
with up to two extra partons in the matrix element, interfaced to
\PYTHIA 8.186~\cite{Sjostrand:2007gs} for parton showers and hadronization. 
The CKKW-L matching scheme~\cite{CKKW} was applied for the matching of 
the matrix element and the parton shower, with a scale parameter set to a quarter of the mass of the sparticle produced. 
The ATLAS A14~\cite{ATL-PHYS-PUB-2014-021} set of tuned parameters (tune) was used for the shower and the underlying event,
 together with the NNPDF2.3~LO~\cite{Ball:2012cx} parton distribution function (PDF) set. 
The {\sc EvtGen} 1.2.0 program~\cite{EvtGen} was used to describe the properties
of the bottom and charm hadron decays in the signal samples.

The signal cross-sections were calculated at next-to-leading order (NLO) in the
strong coupling constant, adding the resummation of soft gluon emission at
next-to-leading-logarithmic accuracy
(NLL)~\cite{Beenakker:1996ch,Kulesza:2008jb,Kulesza:2009kq,Beenakker:2009ha,Beenakker:2011fu}.
The nominal cross-section and its uncertainty are taken from an envelope of
cross-section predictions using different PDF sets and factorization and
renormalization scales, 
as described in Ref.~\cite{Borschensky:2014cia}, 
considering only the four light-flavor left-handed squarks ($\tilde{u}_L$,
$\tilde{d}_L$, $\tilde{s}_L$, and $\tilde{c}_L$).

\begin{table*}[t]
\begin{center}
\caption{Simulated signal and background event samples:
         the corresponding event generator, parton shower, cross-section normalization,
	 PDF set and underlying-event tune are shown. 
}\label{tab:MC}
\footnotesize
\begin{tabular}{llcccc}
\toprule
Physics process & Generator& Parton shower & Cross-section & PDF set & Tune\\
\midrule
Signal & \madgraph 2.2.2 & \pythia8.186 & NLO+NLL & NNPDF2.3 LO & ATLAS A14 \\
\midrule
$\ttbar$         & \textsc{Powheg-Box} v2 & \pythia6.428 & NNLO+NNLL & CT10 NLO
&\textsc{Perugia2012}\\
Single-top       &&&&\\
\quad$t$-channel & \textsc{Powheg-Box} v1 & \pythia6.428 & NLO & CT10f4 NLO &
\textsc{Perugia2012}\\
\quad$s$-channel & \textsc{Powheg-Box} v2 & \pythia6.428 & NLO & CT10 NLO &
\textsc{Perugia2012}\\
\quad$Wt$-channel & \textsc{Powheg-Box} v2 & \pythia6.428 & NLO+NNLL & CT10 NLO
& \textsc{Perugia2012}\\
$W(\rightarrow \ell\nu)$ + jets & \sherpa 2.2.1 & \sherpa& NNLO  &
NNPDF3.0 NNLO & \sherpa default \\
$Z/\gamma^{*}(\rightarrow \ell \ell)$ + jets & \sherpa 2.2.1 & \sherpa &
NNLO & NNPDF3.0 NNLO & \sherpa default\\
\multirow{2}{*}{$WW$, $WZ$ and $ZZ$}   & \sherpa 2.1.1 /
&\multirow{2}{*}{\sherpa}  &\multirow{2}{*}{NLO} & CT10 NLO /  &
\multirow{2}{*}{\sherpa default} \\
&\sherpa 2.2.1&&& NNPDF3.0 NNLO&\\
$\ttbar+W/Z/WW$       &  \madgraph 2.2.2 & \pythia8.186 & NLO  & NNPDF2.3 LO & ATLAS A14 \\
\bottomrule
\end{tabular}
\end{center}
\end{table*}

The simulated event samples for the signal and SM backgrounds are summarized in Table~\ref{tab:MC}.
Additional samples are used to assess systematic uncertainties, as
explained in Section~\ref{sec:systematics}.

To generate \ttbar\ and single-top-quark events in the $Wt$ and
$s$-channel~\cite{ATL-PHYS-PUB-2016-004}, the \textsc{Powheg-Box}
v2~\cite{Alioli:2010xd} event generator with the CT10~\cite{Lai:2010vv} PDF set in the matrix-element calculations was used. Electroweak $t$-channel
single-top-quark events were generated using the \textsc{Powheg-Box} v1 event generator.
This event generator uses the four-flavor scheme for the NLO matrix-element calculations
together with the fixed four-flavor PDF set CT10f4. For all top quark processes,
top quark spin correlations are preserved (for the single-top $t$-channel,
top quarks are decayed using MadSpin~\cite{Artoisenet:2012st}). The parton
shower, fragmentation, and the underlying event were simulated using
\pythia6.428~\cite{Sjostrand:2006za} with the CTEQ6L1~\cite{Pumplin:2002vw} PDF
set and the corresponding \textsc{Perugia2012} tune (P2012)~\cite{Skands:2010ak}. The
top quark mass was set to $172.5 \GeV$. 
The EvtGen 1.2.0 program was also used to describe the properties of the
bottom and charm hadron decays in the \ttbar\ and the single-top-quark samples.
The $h_{\text{damp}}$ parameter, which controls the \pt\ of the first additional
emission beyond the Born configuration, was set to the mass of the top quark.
The main effect of this is to regulate the high-\pt\ emission against which the \ttbar\ system recoils.
The \ttbar\ events are normalized using the cross-sections computed at next-to-next-to-leading order (NNLO) with next-to-next-to-leading-logarithmic (NNLL) corrections~\cite{topxs:2014}.
The single top quark events are normalized using the NLO+NNLL cross-sections for the $Wt$-channel~\cite{wtxs:2014oha}
and to the NLO cross-sections for the $t$- and $s$-channels~\cite{Kant:2014oha}.

Events containing $W$ or $Z$ bosons with associated jets
($W$/$Z$+jets)~\cite{ATL-PHYS-PUB-2016-003} were simulated using the \sherpa2.2.1
event generator~\cite{Gleisberg:2008ta}.
Matrix elements were calculated for up to two partons at NLO and four partons at
LO\@ using the Comix~\cite{Gleisberg:2008fv} and
OpenLoops~\cite{Cascioli:2011va} generators. They were merged with the
\sherpa2.2.1 parton shower~\cite{Schumann:2007mg} with massive $b$- and $c$-quarks using the ME+PS@NLO prescription~\cite{Hoeche:2012yf}.
The NNPDF3.0 NNLO PDF set~\cite{Ball:2014uwa} was used in conjunction with
a dedicated parton shower tuning developed by the \sherpa authors. 
The $W/Z$+jets events are normalized using their NNLO cross-sections~\cite{Gavin:2010az}.

The diboson samples~\cite{ATL-PHYS-PUB-2016-002} were generated using
the \sherpa2.1.1 and 2.2.1 event generators 
using the CT10 and NNPDF3.0 PDF sets, respectively. 
The fully leptonic diboson processes were simulated including final states with four charged leptons, three charged
leptons and one neutrino, two charged leptons and two neutrinos, and one charged lepton and three neutrinos. 
The semileptonic diboson processes were simulated with one of the bosons
decaying hadronically and the other leptonically. The processes were calculated
for up to one parton (for $ZZ$) or no additional partons (for $WW, WZ$) at NLO and up to three partons at LO\@. 

For the $t\bar{t}+W/Z/WW$ processes~\cite{ATL-PHYS-PUB-2016-005}, all events were simulated using \madgraph
2.2.2 at LO interfaced to the \pythia8.186 parton shower
model, with up to two ($t\bar{t}+W$), one ($t\bar{t}+Z$) or no ($t\bar{t}+WW$)
extra partons included in the matrix element. The EvtGen 1.2.0 program~\cite{EvtGen} was used to describe the properties of the
bottom and charm hadron decays. The ATLAS
shower and underlying-event tune A14 was used together with the
NNPDF2.3~LO PDF set. The events are normalized using their
NLO cross-sections~\cite{Lazopoulos:2008,Campbell:2012}.

The response of the detector to particles was modeled either with a full ATLAS
detector simulation \cite{:2010wqa} using
\textsc{Geant4}~\cite{Agostinelli:2002hh} or with a fast
simulation~\cite{atlfast}.
The fast simulation is based on a parameterization of the performance of the
electromagnetic and hadronic calorimeters and on \textsc{Geant4} elsewhere. All
background (signal) samples were prepared using the full (fast) detector
simulation.
All simulated events were generated with a varying number of minimum-bias
interactions overlaid on the hard-scattering event to model the multiple
proton--proton interactions in the same and nearby bunch crossings.
The minimum-bias interactions were simulated with the soft QCD processes of
\pythia8.186 using the A2 tune \cite{ATLAS:2012uec} and the MSTW2008LO PDF
set~\cite{Martin:2009iq}. Corrections were applied to the samples to
account for differences between data and simulation for trigger, identification and reconstruction
efficiencies.

\label{sec:data_trigger}

The proton--proton data analyzed in this paper were collected by ATLAS during 2015 and 2016
 at a center-of-mass energy of 13\,TeV with up to 50 simultaneous interactions
 per proton bunch crossing. 
After application of data-quality
requirements related to the beam and detector conditions, the total integrated
luminosity corresponds to \intlumi.  
The uncertainty in the combined 2015 and 2016 integrated luminosity
is $3.2$\%.
It is derived from a calibration of the luminosity scale using $x$--$y$
beam-separation scans.
This methodology is further detailed in Ref.~\cite{Aaboud:2016hhf}.

The data were collected using the higher-level triggers that select events based
on the magnitude of the missing transverse momentum, \met.
The triggers used are close to fully efficient for events with
an offline-reconstructed \met\ greater than $200\GeV$.

\section{Event reconstruction}
\label{sec:objects}

In each event, proton--proton interaction vertices are reconstructed from at
least two tracks, each with a transverse momentum $\pt > 400 \MeV$ and consistent with the beamspot envelope.
The primary vertex (PV) of the event is selected as the vertex with the largest $\sum\pt^2$ of the associated tracks.

A distinction is made between \emph{preselected} and
\emph{signal} leptons and jets. Preselected leptons and jets are used in the $\met$ computation and are
subject to a series of basic quality requirements. Signal leptons and jets are a subset of the preselected objects with more
stringent requirements and are used for the definition of signal, control and validation regions.

Three-dimensional topological energy clusters in the calorimeters are used as
input to the anti-$k_t$ algorithm with a radius parameter
$R=0.4$~\cite{Cacciari:2008gp,Cacciari:2011ma,ATL-PHYS-PUB-2015-036} to reconstruct preselected
jets. 
The effect of multiple interactions per proton bunch crossing (pileup) is accounted
for using the jet area method~\cite{Cacciari2008119,Aad:2015ina}. 
Subsequent calibrations are applied to the reconstructed jet to improve
the energy resolution~\cite{ATL-PHYS-PUB-2015-015,JES2017}. 
The residual contamination by pileup jets is further suppressed using a
multivariate discriminant that estimates the compatibility of the jet with the PV, as detailed in Ref.~\cite{ATLAS-CONF-2014-018}.

Signal jets must satisfy $\pt>30 \GeV$ and $|\eta|<2.8$. Signal jets within
$|\eta|<2.5$ are identified as candidates for containing $b$-hadrons
($b$-tagged) using the \emph{MV2c10}
algorithm~\cite{ATL-PHYS-PUB-2015-022,ATL-PHYS-PUB-2015-039}. This $b$-tagging
algorithm provides an overall efficiency of $77$\% for jets containing $b$-hadrons in simulated $t\bar{t}$ events, with rejection factors of 6 and 134 on charm and light-jets, respectively~\cite{ATL-PHYS-PUB-2016-012}.

Electron candidates are reconstructed by matching an isolated energy cluster in
the electromagnetic calorimeter to at least one ID track. Preselected electrons
are identified with the likelihood-based \emph{Loose}
criterion described in Ref.~\cite{ATLAS-CONF-2016-024} with additional
requirements on the number of hits in the innermost pixel layer to
discriminate against photon conversions. Furthermore, preselected electrons are required to satisfy
$p_\text{T}>7 \GeV$ and $|\eta|<2.47$.
Muon candidates are formed by a combined re-fitting of tracks reconstructed in
the ID and the MS subsystems. Preselected muons are required to have $\pt>6 \GeV$ and
$|\eta|<2.5$, and satisfy the \emph{Medium} identification criteria in
Ref.~\cite{Aad:2016jkr}.

To avoid double-counting of the preselected jets, electrons, and muons, a sequence of overlap-removal procedures based on the angular distance 
$\Delta R=\sqrt{(\Delta y)^2+(\Delta\phi)^2}$ 
is applied. First, any jet reconstructed within $\Delta R<0.2$ of a preselected
electron is rejected. This prevents electromagnetic energy clusters
simultaneously reconstructed as an electron and a jet from being selected twice.
Next, to remove bremsstrahlung from muons followed by a photon conversion into
electron pairs, electrons within $\Delta R<0.01$ from a preselected muon are
discarded. Subsequently, the contamination from muons from decays of heavy hadrons
is suppressed by removing muons that are within $\Delta R < \min(0.04 +
(10\GeV)/\pt^{\mu}, 0.4)$ from preselected jets meeting the previous criteria,
or $\Delta R<0.2$ from a $b$-tagged jet or a jet containing more than three
tracks with $\pt>500\MeV$. In the former case, the \pt-dependent angular
separation mitigates the rejection of energetic muons close to jets in boosted
event topologies. Finally, jets reconstructed with $\Delta R<0.2$ from a preselected
muon are rejected.

Signal electrons are required to satisfy the likelihood-based \emph{tight}
identification criteria detailed in Ref.~\cite{ATLAS-CONF-2016-024}. Signal muons and electrons satisfy a sequence of $\eta$- and $\pt$-dependent isolation requirements on tracking-based and calorimeter-based variables, defined as the \emph{GradientLoose}~\cite{Aad:2016jkr} isolation criteria. Compatibility of the signal lepton tracks with the PV is enforced by requiring the distance $|z_0\sin\theta|$ to be less than $0.5~\mathrm{mm}$, where $z_0$ is the longitudinal impact parameter. In addition, the transverse impact parameter, $d_0$, divided by its uncertainty, $\sigma(d_0)$, must satisfy $|d_0/\sigma(d_0)|<3$ for signal muons and $|d_0/\sigma(d_0)|<5$ for signal electrons.

Corrections derived from data control samples are applied to simulated events to calibrate the reconstruction and identification efficiencies, the momentum scale and resolution of leptons and the efficiency and mis-tag rate of $b$-tagged jets.

\section{Event selection}
\label{sec:selection}

Each event must satisfy the trigger selection criteria, and must contain a
reconstructed primary vertex.
Non-collision background and detector noise are suppressed by rejecting
events with any preselected jet not satisfying a set of quality
criteria~\cite{ATLAS-CONF-2015-029}.
Exactly one signal lepton, either an electron or a muon, is required.
Events with additional preselected leptons are rejected
to suppress the dilepton \ttbar, single-top ($Wt$-channel), $Z$+jets and diboson
backgrounds. The following observables are used in the definition of signal regions in the analysis.

The missing transverse momentum, \met, is defined as the magnitude of \mpt, the
negative vectorial sum of the transverse momenta of preselected muons,
electrons, jets, and identified and calibrated photons. The calculation of \mpt\
also includes the transverse momenta of all tracks originating from the PV and
not associated with any identified
object~\cite{ATL-PHYS-PUB-2015-023,ATL-PHYS-PUB-2015-027}.

The transverse mass, \mt, is defined from the lepton transverse momentum ${\boldsymbol p}_\mathrm{T}^{\ell}$ and \mpt\ as
\begin{linenomath}
\begin{equation}
    m_{\mathrm{T}} =
\sqrt{2 p_{\mathrm{T}}^{\ell} \met
  (1-\cos[\Delta\phi({\boldsymbol p}_{\mathrm{T}}^{\ell},\mpt)])},
  \nonumber
\end{equation}
\end{linenomath}
where $\Delta\phi({\boldsymbol p}_{\mathrm{T}}^{\ell},\mpt)$ is the azimuthal angle
between ${\boldsymbol p}_{\mathrm{T}}^{\ell}$ and \mpt.
For $W$+jets and semileptonic \ttbar\ events, in which one on-shell $W$ boson decays
leptonically, the observable has an upper endpoint at the $W$-boson mass. The
\mt\ distribution for signal events extends significantly beyond the distributions of the $W$+jets and semileptonic \ttbar\ events.

The effective mass, \meff, is the scalar sum of
the \pt\ of the signal lepton and all signal jets and \met:
\begin{linenomath}
\begin{equation}
    \meff = \ptl + \sum_{j=1}^{N_\mathrm{jet}}p_{\mathrm{T},j} + \met.
      \nonumber
\end{equation}
\end{linenomath}

The effective mass provides good discrimination against SM backgrounds, 
especially for the signal scenarios where energetic jets are expected. 
Gluino production leads to higher jet multiplicity than squark production.
High-mass sparticles tend to produce harder jets than low-mass sparticles.
Thus the optimal \meff\ value depends on the different signal scenarios.
To achieve sensitivity to a wide range of SUSY scenarios with a limited number 
of signal regions, this variable is binned in the final region definition
instead of one simple \meff\ cut.
The detailed description can be found in Section ~\ref{subsec:sr}.

The transverse momentum scalar sum, \HT, is defined as
\begin{linenomath}
\begin{equation}
  \HT =  \sum_{j=1}^{N_\mathrm{jet}}p_{\mathrm{T},j},
    \nonumber
\end{equation}
\end{linenomath}
where the index $j$ runs over all the signal jets in the event.
Empirically, the experimental resolution of \met\ scales with $\sqrt{\HT}$,
and the ratio $\met/\!\sqrt{\HT}$ is useful for suppressing background events
with large \met\ due to jet mismeasurement.

The aplanarity is a variable designed to provide more global information about
the full momentum tensor of the event.
It is defined as $(3/2) \times \lambda_{3}$,
where $\lambda_{3}$ is the smallest eigenvalue of the normalized momentum tensor~\cite{aplanarity:2012} calculated using the momenta of the jets and leptons in the event.
Typical measured aplanarity values lie in the range $0$ -- $0.3$, 
with values near zero indicating relatively planar background-like events.
Signal events tend to have high aplanarity values, since they are more spherical
than background events due to multiple objects emitted in the sparticles decay chains.

\subsection{Signal region definitions}
\label{subsec:sr}

Five sets of event selection criteria, each defining a signal region (SR),
are designed to maximize the signal sensitivity.
Each SR is labeled by the minimum required number of jets and,
optionally, the characteristics of the targeted supersymmetric mass spectrum.
Four of the five SRs, 
\textbf{2J}, \textbf{4J high-x}, \textbf{4J low-x}, and \textbf{6J},
target the gluino/squark one-step models.
The fifth SR,
\textbf{9J},
targets the gluino two-step and pMSSM models.

Table~\ref{tab:SR1} summarizes the four SRs targeting the gluino/squark one-step
models.
The four SRs are mutually exclusive.
For setting model-dependent exclusion limits (``excl''), each of the four SRs is
further binned in $b$-veto/$b$-tag and \meff, and a simultaneous fit is
performed across all 28 bins of the four SRs. This choice enhances the
sensitivity to a range of new-physics scenarios with different properties such as the
presence or absence in the final state of jets containing $b$-hadrons, and different mass
separations between the supersymmetric particles. For model-independent limits
and null-hypothesis tests (``disc'' for discovery), the event yield above a minimum value of \meff\ in each SR is used to search for an
excess over the SM background.

The \textbf{2J} SR provides sensitivity to scenarios characterized
by a relatively heavy \ninoone\ and small differences between
\mgluino, \mchino, and \mnino, where most of the decay products tend to have small $\pt$.
Events with one low-\pt\ lepton and at least two jets are selected.
The minimum lepton $\ptl$ is $7\ (6)\gev$ for the electron (muon), and the maximum \pt\ is scaled
with the number of signal jets in the event as $5\gev\times N_{\mathrm{jet}}$ up to
$35\gev$. The maximum $\ptl$ requirement balances background rejection and signal acceptance for 
models with increasing mass splittings, where there are more energetic lepton and jets. 
Stringent requirements on \met\ and on \meff\ enhance the signal sensitivity 
by selecting signal events in which the final-state neutralinos are boosted against
energetic initial-state radiation (ISR) jets.
The SM background is further suppressed by a tight requirement on $\met/\meff$.

The \textbf{4J high-x} SR is optimized for models where \mnino\ is fixed to 60~\gev\ and $x \approx 1$,
i.e., \mchino\ is close to \mgluino.
The $W$ boson produced in the chargino decay is significantly boosted, giving rise to a high-$\pt$ lepton. 
The main characteristics of signal events in this model are large \mt\ values and relatively soft jets emitted from the sparticle decay. 
Tight requirements are placed on \met, \mt, and $\met/\meff$.

The \textbf{4J low-x} SR targets models where \mnino\ is fixed to 60~\gev\ and $x \approx 0$,
 i.e., \mchino\ is close to \mnino. 
The large \mglsq--\mchino\ mass splitting leads to high jet activity, 
where events are expected to have higher \meff\ and larger aplanarity than in the high-$x$ scenarios.  
The $W$ boson tends to be off-shell, leading to small \mt,
and accordingly an upper bound is imposed to keep this region
orthogonal to the \textbf{4J high-x} SR.

The \textbf{6J} SR is optimized for models with $x=1/2$, targeting scenarios with large sparticle mass.
Events with one high-\pt\ lepton and six or more jets are selected.
Requirements on
\mt, \met, \meff, and aplanarity are imposed to reduce the SM background
from \ttbar\ and $W$\,+\,jets production. 
The sensitivity is improved for scenarios with large \mglsq\ and small
\mnino\ by introducing a higher \meff\ bin. 

\begin{table}[tb]
  \small
  \begin{center}
    \caption{Overview of the selection criteria for the signal regions used for gluino/squark one-step models.}
    \begin{tabular}{lcccc}
      \toprule
      \textbf{SR} & \textbf{2J} & \textbf{4J high-x} & \textbf{4J low-x} & \textbf{6J} \\
      \midrule
      $N_{\ell}$  & $=1$   & $=1$   & $=1$   & $=1$ \\
      \multirow{2}{*}{$\ptl$ [\GeV]} & $>7(6)$ for $e$($\mu$) and &
      \multirow{2}{*}{$>35$} & \multirow{2}{*}{$>35$} & \multirow{2}{*}{$>35$} \\
                                    & $<\min(5\cdot N_{\mathrm{jet}},35)$ & & & \\
      \midrule 
      $N_{\mathrm{jet}}$ & $\geq 2$ & 4--5 & 4--5 & $\geq 6$ \\
      \midrule
      \met\ [\GeV] & $> 430$ & $> 300$ & $> 250$ & $> 350$ \\
      \mt\ [\GeV] & $> 100$ & $> 450$ & 150--450 & $> 175$ \\
      Aplanarity & ... & $>0.01$ & $> 0.05$ & $> 0.06$ \\
      $\met/\meff$ & $> 0.25$ & $> 0.25$ & ... & ... \\
      \midrule
      $N_{b\mathrm{-jet}}$ (excl) & \multicolumn{4}{c}{$=0$ for $b$-veto,
      $\ge1$ for $b$-tag}      \\
      \multirow{2}{*}{\meff\ [\GeV] (excl)}  & 3 bins $\in$ [700,1900]  & 2 bins $\in$ [1000,2000]  &  2 bins $\in$ [1300,2000] &  3 bins $\in$ [700,2300]
      \\
                                             & + [> 1900]           & + [> 2000]             & + [> 2000] & + [> 2300]
                                             \\
      \multirow{2}{*}{\meff\ [\GeV] (disc)}  & \multirow{2}{*}{$>1100$} & \multirow{2}{*}{$>1500$}  &  $> 1650 (1300)$ & $> 2300 (1233)$  \\
                                             & & & for gluino (squark) & for gluino (squark)  \\
      \bottomrule
    \end{tabular}
    \label{tab:SR1}
  \end{center}
\end{table}

Finally, one signal region, \textbf{9J} SR,
is defined to target the pMSSM and gluino two-step models.
The selection criteria are summarized in Table~\ref{tab:SR2}.
At least nine jets are required, targeting the models' long decay chains in which multiple vector or Higgs bosons are produced.
The background is further suppressed by tight requirements on the
aplanarity and on $\met/\!\sqrt{\HT}$.
For setting model-dependent exclusion limits (``excl''),
the SR is separated into $1000<\meff<1500\gev$ and $\meff>1500\gev$
to achieve good discrimination power 
for different gluino masses.
For model-independent null-hypothesis tests (``disc''),
events selected with $\meff>1500\gev$ are used to search for
an excess over the SM background.

\begin{table}[tb]
  \small
  \begin{center}
    \caption{Overview of the selection criteria for the signal region used for pMSSM and gluino two-step models.}
    \begin{tabular}{lc}      
      \toprule
      \textbf{SR}        &\textbf{9J} \\
      \midrule
      $N_{\ell}$         & $= 1$ \\
      $\ptl$  [\GeV]     & $> 35$ \\
      \midrule
      $N_{\mathrm{jet}}$ & $\geq 9$ \\
      \midrule
      \met\ [\GeV]       & $> 200$ \\
      \mt\ [\GeV]        & $> 175$ \\
      Aplanarity         & $> 0.07$ \\
      $\met/\!\sqrt{\HT}$ [\GeV$^{1/2}$] & $\geq 8$ \\
      \meff\ [\GeV] (excl) & $[1000,1500]$, [>1500]  \\
      \meff\ [\GeV] (disc) & $>1500$  \\
      \bottomrule
    \end{tabular}
    \label{tab:SR2}
  \end{center}
\end{table}

\section{Background estimation}
\label{sec:background}

The dominant SM backgrounds in most signal regions originate from top
quark (\ttbar\ and single top) and $W$+jets production. In this section, the techniques employed to estimate the contribution of these 
backgrounds in the signal regions are detailed.

Additional sources of background in all signal regions originate from the production 
of $Z$+jets, $t\bar{t}$ in association with a $W$ or $Z$
boson, and diboson ($WW$, $WZ$, $ZZ$) events. Their contributions are
estimated entirely using simulated event samples normalized to NLO cross-sections.

The contribution from multi-jet processes with a misidentified lepton is found
to be negligible once the lepton isolation and \met\
requirements used in this search are imposed. A data-driven matrix method,
following the implementation described in Ref.~\cite{Aad:2015mia}, determined this
in previous iterations of the analysis~\cite{Aad:2016qqk}.
As this background is found to be negligible, it is not
further considered in the analysis.

The dominant top quark and $W$+jets backgrounds in the \textbf{2J}, \textbf{4J high-x},
\textbf{4J low-x}, and \textbf{6J} signal regions are estimated by
simultaneously normalizing the predicted event yields from simulation to the
number of data events observed in dedicated control regions (CR) using the
fitting procedure described in Section~\ref{sec:result}. The simulation is
then used to extrapolate the measured background rates to the corresponding signal
regions.

The CRs are designed to have high purity in the background process of
interest, a sufficiently large number of events to obtain small
statistical uncertainties in the background prediction, and a small
contamination by events from the signal models under consideration. Moreover,
they are designed to have kinematic properties resembling as closely as possible those of the signal
regions,  in order to provide good estimates of the kinematics of background
processes there. This procedure limits the impact of potentially large
systematic uncertainties in the expected yields from the extrapolation.

Tables~\ref{tab:CRVR2J}--\ref{tab:CRVR6J} list the criteria that define the
control regions corresponding to signal regions \textbf{2J}, \textbf{4J high-x},
\textbf{4J low-x}, and \textbf{6J}. As described in
Section~\ref{sec:selection}, these signal regions contain 
multiple bins in \meff. The same binning is maintained for the
control regions, so that every signal region bin in \meff\ has corresponding
control regions with the same requirements on \meff\ and, therefore, the
backgrounds are estimated independently in each \meff\ bin.

Dedicated top and $W$+jets control regions, respectively denoted by TR and WR,
are constructed in each bin of \meff. The TR and
WR are distinguished by requiring at least one or exactly zero $b$-tagged signal
jets, respectively. Cross-contamination from top and $W$+jets
processes between these two types of control regions is accounted for in the fit. The measured
top and $W$+jets background rates from the TR and WR regions in a given
\meff\ bin are extrapolated to the signal region
within the same \meff\ bin. The signal regions in
a given \meff\ bin may be further separated into
regions with at least one or exactly zero $b$-tagged signal jets as described in
Section~\ref{sec:selection}. For such signal regions separated by $b$-tagged jet
multiplicity, the extrapolation is performed from both the TR and WR regions to each
individual bin of $b$-tagged jet multiplicity.

To validate the extrapolation from control to signal regions using
simulated event samples, dedicated validation regions (VRs) are defined for each set
of control and signal regions. The selection criteria defining these VRs
are also shown in Tables~\ref{tab:CRVR2J}--\ref{tab:CRVR6J}. The same
binning in \meff\ used in the control and signal
regions is also maintained in the validation regions. The VRs are
designed to be kinematically close to the signal regions, with only a small
contamination from the signal in the models considered in this search. The
VRs are not used to constrain parameters in the fit, but provide
a statistically independent cross-check of the extrapolation. The observed event
yields in the VRs are found to be consistent with the background
prediction as further discussed in Section~\ref{sec:result}.

\begin{table}
  \small
  \begin{center}
      \caption{Overview of the control and validation region selection criteria
      corresponding to the \textbf{2J} SR. The top and $W$+jets control regions
      are denoted by TR and WR, respectively.}
    \begin{tabular}{lccc}
      \toprule
       \textbf{2J} & WR / TR & VR \met\ & VR \mt\   \\
      \midrule
      $N_{\mathrm{\ell}}$                & \multicolumn{3}{c}{$=1$} \\
      $\ptl$ [\GeV]                                    & \multicolumn{3}{c}{$>7(6)$ for $e(\mu)$ and $<\min(5\cdot N_{\mathrm{jet}},35)$} \\
      \hline
      $N_{\mathrm{jet}}$                               & \multicolumn{3}{c}{$\geq 2$}      \\
      $N_{b\mathrm{-jet}}$                             & $=0$ / $\geq 1$ & ... & ...    \\
      \hline
      \mt\ [\GeV]                                      & $[40,100]$       & $[40,100]$ & $> 100$ \\
      \met\ [\GeV]                                     & $[300,430]$      & $>430$        & $[300,430]$ \\
      Aplanarity                                & ... & ... & ... \\
      $\met/\meff$           & $> 0.15$ & $> 0.25$ & $> 0.1$  \\
      \meff\ [\GeV]         & \multicolumn{3}{c}{3 bins $\in$ [700,1900] +  [>
      1900]} \\
     \bottomrule
    \end{tabular}
    \label{tab:CRVR2J}
  \end{center}
\end{table}

\begin{table}
  \small
  \begin{center}
      \caption{Overview of the control and validation region selection criteria
      corresponding to the \textbf{4J high-x} SR. The top
      and $W$+jets control regions are denoted by TR and WR, respectively.}
    \begin{tabular}{lcccc}
      \toprule
       \textbf{4J high-x} & WR / TR & VR Aplanarity & VR \mt\ & VR Hybrid  \\
      \midrule
      $N_{\mathrm{\ell}}$                & \multicolumn{4}{c}{$=1$} \\
      $\ptl$ [\GeV]                                    & \multicolumn{4}{c}{$>35$} \\
      \hline
      $N_{\mathrm{jet}}$                               & \multicolumn{4}{c}{4--5} \\
      $N_{b\mathrm{-jet}}$                             & $=0$ / $\geq 1$ & ... & ... & ... \\
      \hline
      \mt\ [\GeV]                                      & $[50,200]$ & $[50,150]$ & $>200$ & $[150,450]$ \\
      \met\ [\GeV]                                     & $>300$      & $>250$ & $>250$ & $>250$ \\
      Aplanarity                                & $< 0.01$     &$> 0.05$ & $<0.01$ & $[0.01,0.05]$ \\
      $\met/\meff$           & $>0.25$ & $>0.25$  & $> 0.25$ & ... \\
      \meff\ [\GeV]         & \multicolumn{4}{c}{2 bins $\in$ [1000,2000] + [>
      2000]}
      \\
     \bottomrule
    \end{tabular}
    \label{tab:CRVR4Jhighx}
  \end{center}
\end{table}

\begin{table}
  \small
  \begin{center}
      \caption{Overview of the control and validation region selection criteria
      corresponding to the \textbf{4J low-x} SR. The top and
      $W$+jets control regions are denoted by TR and WR, respectively.}
    \begin{tabular}{lccc}
      \toprule
       \textbf{4J low-x} & WR / TR & VR Aplanarity & VR Hybrid  \\
      \midrule
      $N_{\mathrm{\ell}}$                & \multicolumn{3}{c}{$=1$} \\
      $\ptl$ [\GeV]                                    & \multicolumn{3}{c}{$>35$} \\
      \hline
      $N_{\mathrm{jet}}$                               & \multicolumn{3}{c}{4--5} \\
      $N_{b\mathrm{-jet}}$                             & $=0$ / $\geq 1$ & ... & ... \\
      \hline
      \mt\ [\GeV]                                      & $[50,150]$ & $[50,150]$ & $[150,450]$ \\
      \met\ [\GeV]                                     & \multicolumn{3}{c}{$>250$} \\
      Aplanarity                                & $[0.01,0.05]$ &$>0.05$ & $[0.01,0.05]$ \\
      \meff\ [\GeV]         & \multicolumn{3}{c}{2 bins $\in$ [1300,2000] + [>
      2000]}
      \\
     \bottomrule
    \end{tabular}
    \label{tab:CRVR4Jlowx}
  \end{center}
\end{table}

\begin{table}
  \small
  \begin{center}
      \caption{Overview of the control and validation region selection criteria
      corresponding to the \textbf{6J} SR. The top
      and $W$+jets control regions are denoted by TR and WR, respectively.}
    \begin{tabular}{lccc}
      \toprule
       \textbf{6J} & WR / TR & VR Aplanarity & VR \mt\ \\
      \midrule
      $N_{\mathrm{\ell}}$                 & \multicolumn{3}{c}{$=1$} \\
      $\ptl$ [\GeV]                                    & \multicolumn{3}{c}{$>35$} \\
      \hline
      $N_{\mathrm{jet}}$                               &\multicolumn{3}{c}{$\geq6$} \\
      $N_{b\mathrm{-jet}}$                             & $=0$ / $\geq 1$ & ... & ... \\
      \hline
      \mt\ [\GeV]                                      & $[50,175]$ & $[50,175]$ & $[175,400]$ \\
      \met\ [\GeV]                                     & $>350$      & $>350$ &  $>250$  \\
      Aplanarity                                & $< 0.06$     &$> 0.06$ & $< 0.06$ \\
      \meff\ [\GeV]         & \multicolumn{3}{c}{3 bins $\in$ [700,2300] + [>
      2300]} \\
     \bottomrule
    \end{tabular}
    \label{tab:CRVR6J}
  \end{center}
\end{table}

One of the dominant background components in the \textbf{2J}, \textbf{4J high-x}, 
\textbf{4J low-x}, and \textbf{6J} SRs
is \ttbar\ production with
dileptonic final state, where one lepton fails to be reconstructed (``missing
lepton'') or is a semi-hadronically decaying $\tau$ lepton; this background is
characterized by high values of \mt.
To validate the above
described background estimation technique, which is largely a simulation-based extrapolation from low-\mt\ control regions populated by events with
semileptonic \ttbar\ decays, an alternative method was developed. This
method (hereafter referred to as the object replacement method) uses events
in a dileptonic control region.
To emulate the missing lepton case, the \pt\ of one of the two leptons is added
vectorially to the calculation of \met.
To emulate the hadronic $\tau$ decay case, one of the two leptons is
re-simulated as a hadronic tau decay using the 
Tauola generator~\cite{Was:2011tv} with appropriate energy scale and resolution
corrections. The accuracy of this alternative background estimation technique
was validated on simulated samples as well as in data
validation regions. The background estimates derived from this object
replacement method are found to be consistent with those
obtained from the standard semi-data-driven approach as further demonstrated in
Section~\ref{sec:result}.

\begin{figure}
  \begin{center}
    \includegraphics[width=0.6\textwidth]{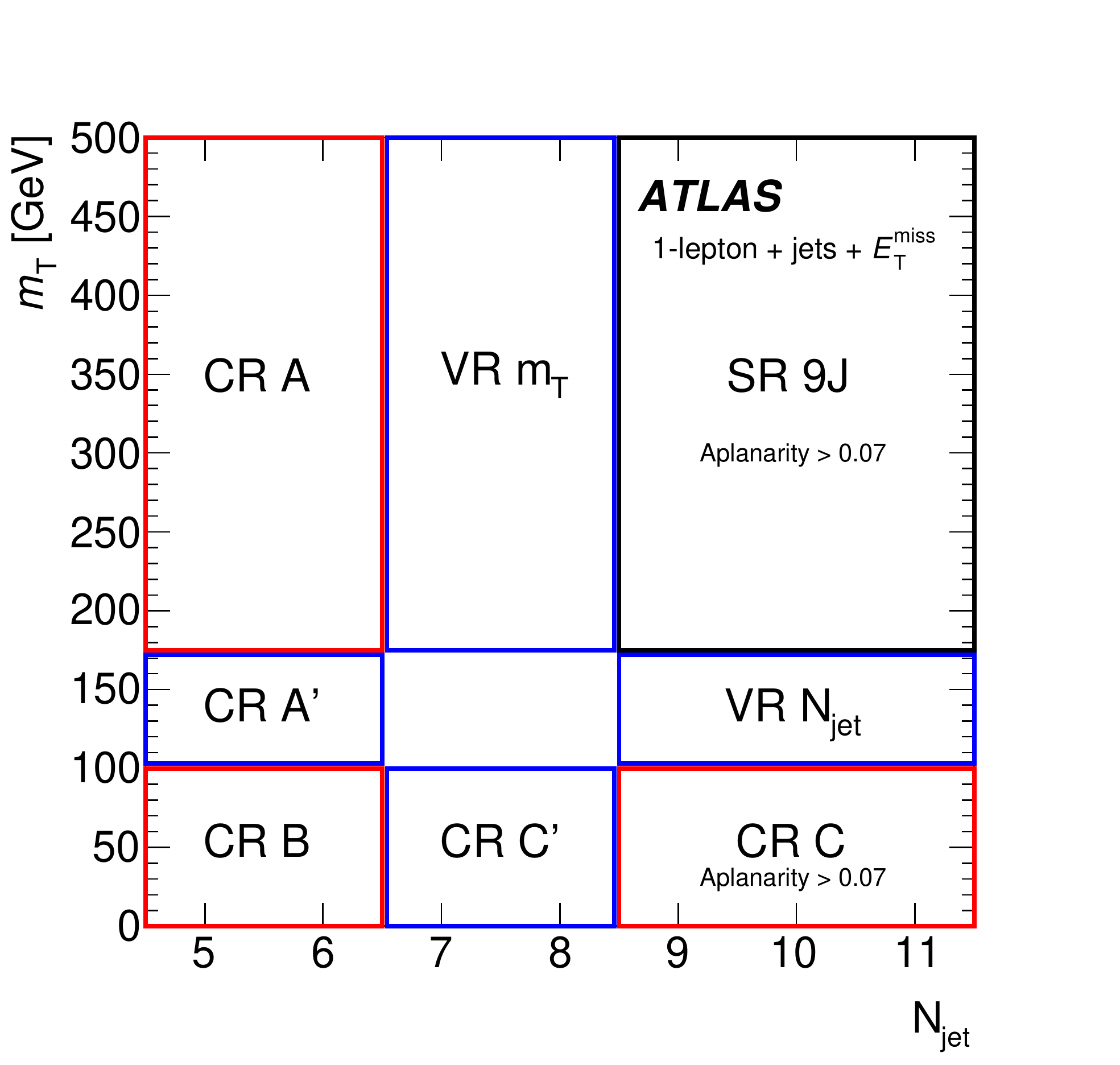}
  \end{center}
  \caption{Illustration of the control and validation region
    configuration corresponding to the \textbf{9J} SR. The control regions
    that are used for the background estimation in the signal region are
    indicated by red lines, while the blue lines indicate control regions
    that are used for the background estimation in the validation regions
    and the validation regions themselves.  The definitions for all
    regions are given in Table~\ref{tab:SR2} (signal region) and
    Table~\ref{tab:CRVR9J} (control and validation regions).}
  \label{fig:multijet-regions}
\end{figure}

While the background estimation strategy described above works well
for the signal regions \textbf{2J}, \textbf{4J high-x}, \textbf{4J low-x}, and \textbf{6J},
it is not viable for the \textbf{9J} SR.
The reason for this is that the simulation-based
extrapolation from the control regions, which are typically located around the
peak region of the transverse mass distribution  ($\mt \sim 80 \gev$), to the
high-\mt\ signal regions  ($\mt \gg 80 \gev$) is affected by large theoretical
uncertainties at high jet multiplicities.
Because the peak and tail regions of the \mt\ distribution are dominated by
semileptonic and dileptonic final states from \ttbar\ decays, respectively, additional jets from initial- 
or final-state radiation are required to obtain the same jet multiplicity for
dileptonic \ttbar\ final states.
Inadequate modeling of such additional jets is the dominant source of the
theoretical uncertainty.
To reduce the dependence on the modeling of additional jets, a dedicated
data-driven background estimation technique was designed for the 
\textbf{9J} SR. The method relies on the assumption that the \mt\ distribution
is approximately invariant under changes in the jet multiplicity requirements.
This assumption is found to be valid when tight \meff\ requirements as used in
this analysis are applied such that the overall activity in the calorimeter and
thus the missing transverse momentum resolution are not significantly affected
by variations in the jet multiplicity. Based on the \mt\ invariance, mutually
exclusive control regions CR$_{\mathrm{A,B,C}}$ are defined in the
\mt--$N_{\mathrm{jet}}$ plane, where CR$_{\mathrm{A}}$ is located at high \mt\
and low $N_{\mathrm{jet}}$, CR$_{\mathrm{B}}$ at low \mt\ and
low $N_{\mathrm{jet}}$, and CR$_{\mathrm{C}}$ at low \mt\ and
high $N_{\mathrm{jet}}$. The precise requirements of these regions are defined
in Table~\ref{tab:CRVR9J} and illustrated in Figure~\ref{fig:multijet-regions}.
Based on these regions, the background in the high \mt~and high
$N_{\mathrm{jet}}$ signal region can then be estimated with the following equation

\begin{table}
  \small
  \begin{center}
      \caption{Overview of the control and validation region selection criteria
      corresponding to the \textbf{9J} SR. The control regions
      CR$_{\mathrm{A,A',B,C,C'}}$ are further divided into bins of exactly $0$
      or $\geq 1$ $b$-tagged signal jets to enrich top and $W$+jets backgrounds,
      respectively.}
	\begin{tabular}{lccccccc}
    \toprule	
    \textbf{9J}                           & CR$_\mathrm{A}$ & CR$_\mathrm{B}$ & CR$_\mathrm{C}$  & VR \mt\ & CR$_\mathrm{C'}$ & VR $N_{\mathrm{jet}}$ & CR$_\mathrm{A'}$ \\
    \midrule
    $N_{\ell}$                    & \multicolumn{7}{c}{$= 1$}           \\
    \ptl\ [\GeV]                          & \multicolumn{7}{c}{$\ge 35$}         \\
    \hline
    $N_{\mathrm{jet}}$                    & 5--6 & 5--6 & $\ge 9$ & 7--8 & 7--8 & $\ge 9$        & 5--6 \\
    $p_{\mathrm{T}}^{\mathrm{jet}}$ [\GeV] & \multicolumn{7}{c}{$\ge 30$}                                      \\
    $N_{b\mathrm{-jet}}$                  & $=0$ / $\geq 1$ & $=0$ / $\geq 1$ &
    $=0$ / $\geq 1$ & ... & $=0$ / $\geq 1$ & ... & $=0$ / $\geq 1$ \\
	\hline
    \mt\ [\GeV]                           & $> 175$ & $<100$ & $<100$ & $> 175$  & $< 100$    & $[100,175]$ & $[100,175]$ \\
    $\met$ [\GeV]                          & \multicolumn{7}{c}{$> 200$}                                       \\
    $\met/\sqrt{H_{\mathrm{T}}}$               & \multicolumn{7}{c}{$\ge 8$}        \\
    Aplanarity                     & ... & ... & $> 0.07$ & $<0.05$ & $<0.05$ & $<0.05$ & ...     \\
    \multirow{2}{*}{$\meff$ [\GeV]} & \multirow{2}{*}{$>1000$} &
    \multirow{2}{*}{$>1000$} & \multirow{2}{*}{$>1000$} & $[1000, 1500]$, & \multirow{2}{*}{$> 1000$} & \multirow{2}{*}{$> 1000$} & \multirow{2}{*}{$> 1000$} \\
    \multirow{2}{*}{}                                         &
    \multicolumn{3}{c}{\multirow{2}{*}{}}         & [> 1500] &
    \multicolumn{3}{c}{\multirow{2}{*}{}}        \\
    \bottomrule
	\end{tabular}
    \label{tab:CRVR9J}
  \end{center}
\end{table}
\begin{linenomath}
\begin{equation}
\frac{N_{\mathrm{CR_A}}}{N_{\mathrm{CR_B}}}=\frac{N_{\mathrm{SR_{9J}}}^{\mathrm{est}}}{N_{\mathrm{CR_C}}}\quad\rightarrow\quad
N_{\mathrm{SR_{9J}}}^{\mathrm{est}}=\frac{N_{\mathrm{CR_A}}}{N_{\mathrm{CR_B}}}N_{\mathrm{CR_C}},
  \nonumber
\end{equation}
\end{linenomath}
where $N_{\mathrm{<region>}}^{\mathrm{est}}$ is the (estimated) number of
events in a given region. The residual small correlations between \mt\ and $N_{\mathrm{jet}}$ that bias
the background estimate in the signal region can then be expressed in terms of a
simulation-based closure parameter defined as
\begin{linenomath}
\begin{equation}
f_{\mathrm{closure}} =
\frac{N_{\mathrm{SR_{9J}}}^{\mathrm{sim}}}{N_{\mathrm{SR_{9J}}}^{\mathrm{sim,est}}} = \frac{N_{\mathrm{SR_{9J}}}^{\mathrm{sim}}\cdot N_{\mathrm{CR_B}}^{\mathrm{sim}}}{N_{\mathrm{CR_C}}^{\mathrm{sim}} \cdot N_{\mathrm{CR_A}}^{\mathrm{sim}}},
  \nonumber
\end{equation}
\end{linenomath}

where $N_{\mathrm{<region>}}^{\mathrm{sim}}$ is the number of
events in a given region as predicted by simulation while 
$N_{\mathrm{SR_{9J}}}^{\mathrm{sim,est}}$ is the estimated number of events in
the signal region based on the simulation predictions in regions A, B, and C.
The estimated number of background events in the signal region can then be rewritten as
\begin{linenomath}
\begin{eqnarray}
N_{\mathrm{SR_{9J}}}^{\mathrm{est}}= f_{\mathrm{closure}}\cdot\frac{N_{\mathrm{CR_A}}^{\mathrm{obs}}}{N_{\mathrm{CR_B}}^{\mathrm{obs}}} \cdot N_{\mathrm{CR_C}}^{\mathrm{obs}} &=&
  \nonumber
N_{\mathrm{SR_{9J}}}^{\mathrm{sim}} \cdot
\left(\frac{N_{\mathrm{CR_A}}}{N_{\mathrm{CR_B}}}N_{\mathrm{CR_C}}\right)^{\mathrm{obs}}\bigg/
\left(\frac{N_{\mathrm{CR_A}}}{N_{\mathrm{CR_B}}}N_{\mathrm{CR_C}}\right)^{\mathrm{sim}}\\
 &=&
N_{\mathrm{SR_{9J}}}^{\mathrm{sim}}\cdot\underbrace{\frac{N_{\mathrm{CR_C}}^{\mathrm{obs}}}{N_{\mathrm{CR_C}}^{\mathrm{sim}}}}_{\mu_\mathrm{C}}\cdot\underbrace{\frac{\left(\frac{N_{\mathrm{CR_A}}}{N_{\mathrm{CR_B}}}\right)^\mathrm{obs}}{\left(\frac{N_{\mathrm{CR_A}}}{N_{\mathrm{CR_B}}}\right)^\mathrm{sim}}}_{\mu_\mathrm{A/B}},
  \nonumber
\end{eqnarray}
\end{linenomath}
where $N_{\mathrm{<region>}}^{\mathrm{obs}}$ is the observed number of events in
a given region, $\mu_\mathrm{C}$ is the normalization parameter in region C, and
the normalization parameter $\mu_\mathrm{A/B}$ is fitted simultaneously with the normalization $\mu_\mathrm{B}$ of the backgrounds in region CR$_{\mathrm{
B}}$ according to
\begin{linenomath}
\begin{equation}
N_{\mathrm{CR_B}}^\mathrm{obs} =
N_{\mathrm{CR_B}}^\mathrm{sim}\cdot\underbrace{\frac{N_{\mathrm{CR_B}}^\mathrm{obs}}{N_{\mathrm{CR_B}}^\mathrm{sim}}}_{\mu_\mathrm{B}}\quad {\mathrm{and}}\quad N_{\mathrm{CR_A}}^\mathrm{obs} = N_{\mathrm{CR_A}}^\mathrm{sim}\cdot\underbrace{\frac{\left(\frac{N_{\mathrm{CR_A}}}{N_{\mathrm{CR_B}}}\right)^\mathrm{obs}}{\left(\frac{N_{\mathrm{CR_A}}}{N_{\mathrm{CR_B}}}\right)^\mathrm{sim}}}_{\mu_\mathrm{A/B}}\underbrace{\frac{N_{\mathrm{CR_B}}^\mathrm{obs}}{N_{\mathrm{CR_B}}^\mathrm{sim}}}_{\mu_\mathrm{B}}.
  \nonumber
\end{equation}
\end{linenomath}

The control regions listed in Table~\ref{tab:CRVR9J} are optimized to
provide a sufficient number of events in the backgrounds of interest,
low contamination from the signal models considered, and a closure parameter
$f_{\mathrm{closure}}$ close to unity. All control regions are fitted simultaneously
in two bins requiring either zero or at least one $b$-tagged signal jet to
enrich the contributions from the $W$+jets and top backgrounds, respectively.
Therefore, the normalization factors $\mu_\mathrm{B}$, $\mu_\mathrm{C}$, and
$\mu_\mathrm{A/B}$ exist separately for the $W$+jets and top backgrounds. The
top backgrounds considered in the fit comprise \ttbar\ as well as single-top
production processes, which are treated with a common
set of normalization parameters.

To validate that the fitted ratio of low-\mt{} to high-\mt{} events
($\mu_\mathrm{A/B}$) extrapolates to high values of $N_\mathrm{jet}$, a
validation region VR \mt\ with seven or eight jets and high \mt{} requirements is introduced.
Similarly, a validation region \mbox{VR $N_{\mathrm{jet}}$} with at least nine
jets and moderate \mt{} requirements is introduced to validate the extrapolation of
the normalization factor $\mu_{\mathrm{C}}$ in region CR$_{\mathrm{C}}$ to higher \mt{}
values. Since the normalization factors for different jet multiplicities are
expected to differ, a control region CR$_{\mathrm{C'}}$ along with its normalization
factor ($\mu_\mathrm{C'}$) is introduced. This region is only used to obtain the
background estimate in VR \mt. Similarly, a control region CR$_{\mathrm{A'}}$ is constructed to obtain the normalization factor $\mu_\mathrm{A'/B}$ that
is needed for the background estimation in validation region VR
$N_{\mathrm{jet}}$. The definition of the validation regions along with their
corresponding control regions is given in Table~\ref{tab:CRVR9J}.

\section{Systematic uncertainties}
\label{sec:systematics}

Experimental and theoretical sources of systematic uncertainty are described in this section. Their effects are evaluated for all simulated signal and background events.

The dominant experimental systematic effects are the uncertainties associated
with the jet energy scale (JES) and resolution (JER) and with the $b$-tagging
efficiency and mis-tagging rate. The impact of the jet-related uncertainties on the total
background prediction ranges from 1.3\% in the \textbf{6J} SR to 18\% in the
\textbf{9J} SR. Similarly, the impact of the uncertainties associated with the
$b$-tagging procedure amounts to 1.9\% in the \textbf{6J} SR bins with at
least one $b$-tagged jet and increases to 9.5\% in the \textbf{6J} SR bins
with no $b$-tagged jets.
The simulation is reweighted to match the distribution of the average number of
proton--proton interactions per bunch crossing ($\mu$) observed in data. The
uncertainty in $\mu$ is propagated by varying up and down the reweighting factor: it becomes relevant in the signal regions characterized by the highest jet multiplicities.

Uncertainties in the theoretical predictions and the modeling of simulated
events are also considered. For the $W$+jets and the \ttbar\ and single top
backgrounds, they affect the extrapolation from each $\meff$ bin in the control
regions to the corresponding bin in the signal regions. In the \textbf{9J} SR
the $f_{\text{closure}}$ parameter used in the background estimation in this
channel is affected as well.
For all the other background sources, they impact the inclusive cross-section of each specific process, the acceptance of the analysis selection requirements and the shape of the $\meff$ distribution in each SR.

An uncertainty stems from the choice of MC event generator modeling the
$t\bar{t}$, single top, diboson and $W/Z$+jets processes.
For $t\bar{t}$~ and single top, \textsc{Powheg-Box} is compared with
\textsc{MG5\_aMc@NLO}~\cite{Alwall:2014hca}
and the relative difference in the extrapolation factors is evaluated.
For $W/Z$+jets, the predictions from \sherpa are compared with \textsc{MG5\_aMc@NLO}~\cite{Alwall:2014hca}.
For dibosons, the event yield predictions from \sherpa are compared with \textsc{powheg-box} interfaced to \textsc{Pythia}. The impact of varying
the amount of initial- and final-state radiation is evaluated for $t\bar{t}$ and
single top production. Specific samples are used, with altered renormalization
and factorization scales as well as parton shower and NLO radiation~\cite{ATL-PHYS-PUB-2016-004}. Moreover,
the difference between the predictions from \textsc{powheg-box} interfaced to
\textsc{Pythia} and to \textsc{Herwig++}~\cite{Bahr:2008pv} is
computed to estimate the uncertainty associated with the parton shower modeling.
For $W/Z$+jets samples, the uncertainties in the renormalization, factorization, resummation scales and the matching scale between matrix elements and parton shower (CKKW-L) are evaluated by varying up and down by a factor of two the corresponding parameters in \textsc{Sherpa}. For $t\bar{t}$ and $W$+jets samples, the uncertainties due to choosing the PDF set CT10~\cite{Lai:2010vv} are considered.

Inclusive $WWbb$ events generated using
\textsc{MG5\_aMC@NLO}~\cite{Alwall:2014hca} are compared to the sum of \ttbar\
and $Wt$ production, to assign an uncertainty to the interference effects
between single top and \ttbar\ production at NLO. The uncertainty in the inclusive $Z$+jets cross-section, amounting to 5\%, is accounted for~\cite{ATLAS-CONF-2015-039}. An overall 6\% systematic uncertainty in the inclusive cross-section of diboson processes is also considered. In addition, the \sherpa parameters controlling the renormalization, factorization, resummation and matching scales are varied by a factor of two to estimate the corresponding uncertainties. An uncertainty of $30$\% is assigned to the small contributions of $t\bar{t}+W/Z/WW$.

The total systematic uncertainty in the predicted background yields in the
various signal regions ranges from 12\% in the \textbf{2J} SR bins with $\geq
1$~$b$-tagged jet, to 50\% in the \textbf{9J} SR. The largest uncertainties in the SR
bins with $\geq 1$~$b$-tagged jet originate from the modeling of \ttbar\ events and
amount to 5\% in the \textbf{2J} SR, increasing to 40\% in the \textbf{9J} SR.
Similarly, in the SR bins where $b$-tagged jets are vetoed, the dominant source of
systematic uncertainty is the modeling of $W$+jets events, ranging from 9\% in
the \textbf{6J} SR to 20\% in the \textbf{4J low-x} SR. Other important
uncertainties are those associated with the finite size
of the MC samples, which amount to 18\% in the \textbf{6J} SR, and the theoretical
uncertainties originating from the modeling of the diboson background,
amounting to 26\% in the \textbf{6J} SR. Tables~\ref{syst24J}--\ref{syst9J}
list the breakdown of the dominant systematic uncertainties in background estimates in the various signal regions.
\begin{table}
\begin{center}

\caption[Breakdown of uncertainty on background estimates]{
Breakdown of the dominant systematic uncertainties in the background estimates
in the {\bf 2J} and {\bf 4J high-x} SRs.
The individual uncertainties can be correlated and do not necessarily add up
in quadrature to the total background uncertainty. The percentages show the size
of the uncertainty relative to the total expected background.\label{syst24J}}
\setlength{\tabcolsep}{0.0pc}
\begin{tabular*}{\textwidth}{@{\extracolsep{\fill}}lcccc}
\noalign{\smallskip}\hline\noalign{\smallskip}
\textbf{ Signal region }          &  \textbf{2J} $b$-tag  & \textbf{2J} $b$-veto   & \textbf{4J high-x} $b$-tag   & \textbf{4J high-x} $b$-veto \\
\noalign{\smallskip}\hline\noalign{\smallskip}

Total background expectation             &  $47$        &  $36$     &  $54$        &  $44$   \\

\noalign{\smallskip}\hline\noalign{\smallskip}
Total background systematic uncertainty               & $\pm 4\ [9\%] $        & $\pm 9\ [24\%] $        & $\pm 7\ [12\%] $        & $\pm 10\ [23\%] $ \\
\noalign{\smallskip}\hline\noalign{\smallskip}
\noalign{\smallskip}\hline\noalign{\smallskip}

Experimental uncertainty           & $\pm 1.3$          & $\pm 2.2 $           & $\pm 2.6$          & $\pm 5 $         \\

Normalization uncertainty          & $\pm 2.8 $          & $\pm 0.9 $            & $\pm 4$          & $\pm 1.9$        \\

Theoretical uncertainty           & $\pm 3.5 $          & $\pm 9 $            & $\pm 5 $          & $\pm 6 $  \\

Statistical uncertainty of MC samples         & $\pm 1.4 $          & $\pm 1.8 $
& $\pm 1.7 $          & $\pm 7 $ \\

\noalign{\smallskip}\hline\noalign{\smallskip}
\end{tabular*}
\end{center}
\end{table}

\begin{table}
\begin{center}
\caption[Breakdown of uncertainty on background estimates]{
Breakdown of the dominant systematic uncertainties in the background estimates
in the {\bf 4J low-x} and {\bf 6J} SRs.
The individual uncertainties can be correlated and do not necessarily add up
in quadrature to the total background uncertainty. The percentages show the size of the uncertainty relative to the total expected background.\label{syst46J}}
\setlength{\tabcolsep}{0.0pc}
\begin{tabular*}{\textwidth}{@{\extracolsep{\fill}}lcccc}
\noalign{\smallskip}\hline\noalign{\smallskip}
\textbf{ Signal region }          & \textbf{4J low-x} $b$-tag & \textbf{4J low-x} $b$-veto  & \textbf{6J} $b$-tag      &    \textbf{6J} $b$-veto    \\
\noalign{\smallskip}\hline\noalign{\smallskip}

Total background expectation            &  $31$        &  $16$             &  $27$        &  $7.3$       \\
\noalign{\smallskip}\hline\noalign{\smallskip}

Total background systematic uncertainty                  & $\pm 6\ [21\%] $        & $\pm 4\ [25\%] $     & $\pm 4\ [15\%] $        & $\pm 2.0\ [27\%] $             \\
\noalign{\smallskip}\hline\noalign{\smallskip}
\noalign{\smallskip}\hline\noalign{\smallskip}

Experimental uncertainty         & $\pm 1.8 $          & $\pm 1.0 $          & $\pm 1.1 $          & $\pm 0.8 $       \\

Normalization uncertainty           & $\pm 2.3$          & $\pm 0.8 $              & $\pm 1.4 $          & $\pm 0.5 $       \\

Theoretical uncertainty        & $\pm 6 $          & $\pm 4 $               & $\pm 4 $          & $\pm 2.0 $       \\

Statistical uncertainty of MC samples  & $\pm 1.2$          & $\pm 1.0 $        
& $\pm 0.9 $          & $\pm 0.6 $       \\

\noalign{\smallskip}\hline\noalign{\smallskip}
\end{tabular*}
\end{center}
\end{table}

\begin{table}
\begin{center}
\caption[Breakdown of uncertainty on background estimates]{
Breakdown of the dominant systematic uncertainties in the background estimates
in the {\bf 9J} SR.
The individual uncertainties can be correlated and do not necessarily add up
in quadrature to the total background uncertainty. The percentage shows the size of the uncertainty relative to the total expected background.
\label{syst9J}}
\setlength{\tabcolsep}{0.0pc}
\begin{tabular*}{\textwidth}{@{\extracolsep{\fill}}lc}
\noalign{\smallskip}\hline\noalign{\smallskip}
\textbf{ Signal region }                                    & \textbf{9J}      \\
\noalign{\smallskip}\hline\noalign{\smallskip}

Total background expectation             &  $7$       \\

\noalign{\smallskip}\hline\noalign{\smallskip}

Total background systematic uncertainty               & $\pm 4\ [50\%] $             \\
\noalign{\smallskip}\hline\noalign{\smallskip}
\noalign{\smallskip}\hline\noalign{\smallskip}

Theoretical uncertainty         & $\pm 4$       \\

Normalization uncertainty         & $\pm 2.0$       \\

Experimental uncertainty         & $\pm 1.9$       \\

Statistical uncertainty of MC samples         & $\pm 0.7$       \\

\noalign{\smallskip}\hline\noalign{\smallskip}
\end{tabular*}
\end{center}
\end{table}

For the signal processes, the modeling of initial-state radiation can be
affected by sizable theoretical uncertainty. The uncertainties in the expected
yields for SUSY signal models are estimated with variations of a factor of two to the \textsc{MG5\_aMC@NLO} parameters corresponding to the renormalization, factorization and jet matching scales, and to the \textsc{Pythia} shower tune parameters. The overall uncertainties range from about 1\% for signal models with large mass splitting between the gluino or squark, the chargino, and the neutralino, to 35\% for models with very compressed mass spectra.

\section{Results and Interpretation}
\label{sec:result}

The statistical interpretation of the results is performed 
based on a profile likelihood method~\cite{Cowan:2010js} 
using the HistFitter framework~\cite{Baak:2014wma}.
The likelihood function consists of a product of Poisson probability density functions
for the signal and control regions that contribute to the fit.
The inputs to the likelihood function are the observed numbers of data events
and the expected numbers of signal and SM background events in each region.
Three normalization factors, one for signal, one for $W$\,+\,jets, 
one for \ttbar\  and single top, are introduced
to adjust the relative contributions of the main background and signal components.
The small sources of SM background, i.e.,  diboson, $Z$\,+\,jets and $t\bar t+V$, are estimated directly from simulation.
The uncertainties are implemented in the fit as nuisance parameters,
which are correlated between the SRs and the CRs.
The systematic uncertainties described in Section~\ref{sec:systematics} are
constrained by Gaussian probability density functions, while the statistical uncertainties are constrained by Poisson probability density functions.

The observed numbers of events in the signal regions
are given in Tables~\ref{results:tab:yieldsSRs_2J}--\ref{results:tab:yieldsSRs_9J},
along with the SM background prediction 
as determined with the \textit{background-only} fit.
In a background-only fit, the data event yields in the CRs are used to determine
the two background normalization factors: for $W$\,+\,jets and for \ttbar\ and
single top production. 
The fit is independent of the observation in the SR, and does not consider
signal contamination in the CRs. The above-mentioned signal normalization
parameter is therefore not included in this fit configuration. 

The compatibility of the observed and expected event yields in both the
validation and signal regions is illustrated 
in Figures~\ref{results:2JVRSRPulls}--\ref{results:9JVRSRPulls}. 
No significant excess in data is observed over the SM prediction.

\newcommand{\bgonlycaption}[1]{
Event yields and background-only fit results for the #1.
Each column corresponds to a bin in \meff\ [\GeV].
Uncertainties in the fitted background estimates combine statistical (in the
simulated event yields) and systematic uncertainties. The uncertainties in this
table are symmetrized for propagation purposes but truncated at zero to remain
within the physical boundaries.}

\begin{table}
  \begin{center}
    \caption{\label{results:tab:yieldsSRs_2J}\bgonlycaption{{\bf 2J} and {\bf
    4J high-x} SRs}}

\begin{tabular*}{\textwidth}{@{\extracolsep{\fill}}lccccc}
\toprule
\textbf{2J} $b$-tag & All \meff\ bins & [700,1100] & [1100,1500] & [1500,1900] &
$>1900$ [\GeV] \\
\midrule

Observed events          & $47$              & $8$              & $21$              & $12$              & $6$                    \\
\midrule

Fitted bkg events         & $47 \pm 4$          & $6.0 \pm 1.3$          & $23.0 \pm 3.0$          & $12.9 \pm 2.1$          & $5.2 \pm 1.8$              \\
\midrule

        Fitted \ttbar\ events         & $31.1 \pm 3.5$          & $2.7 \pm  0.8$ & $15.3 \pm 2.5$          & $9.4 \pm 1.7$          & $3.8 \pm 1.4$              \\

        Fitted $W$+jets events         & $3.7 \pm 1.4$          & $1.1 \pm  0.9$ & $2.0 \pm 1.0$          & $0.5 \pm 0.4$          & $0.10 \pm 0.10$              \\

        Fitted $Z$+jets events         & $2.0 \pm 0.6$          & $0.70 \pm  0.20$ & $0.90 \pm 0.30$          & $0.30 \pm 0.10$          & $0.10 \pm 0.10$              \\

        Fitted single-top events         & $5.6 \pm 1.8$          & $0.6 \pm   0.4$          & $2.7 \pm 1.3$          & $1.5 \pm 1.0$          & $0.8 \pm 0.8$              \\

        Fitted diboson events         & $1.8 \pm 1.4$          & $0.30 \pm 0.20$          & $0.8 \pm 0.8$          & $0.50 \pm 0.30$          & $0.10 \pm 0.10$              \\

        Fitted \ttbar\+V events         & $2.92 \pm 0.25$          & $0.60 \pm 0.10$ & $1.30 \pm 0.10$          & $0.70 \pm 0.10$          & $0.30 \pm 0.04$              \\

\toprule
\textbf{2J} $b$-veto & All \meff\ bins & [700,1100] & [1100,1500] & [1500,1900]
& $>1900$ [\GeV] \\
\midrule

Observed events          & $61$              & $20$              & $26$              & $9$              & $6$                    \\
\midrule

Fitted bkg events         & $36 \pm 9$          & $10 \pm 4$          & $16 \pm 5$          & $7.0 \pm 1.9$          & $2.5 \pm 0.8$              \\
\midrule

        Fitted \ttbar\ events         & $5.1 \pm 1.0$          & $1.00 \pm 0.30$       & $2.7 \pm 0.6$          & $1.00 \pm 0.30$          & $0.40 \pm 0.20$              \\

        Fitted $W$+jets events         & $13 \pm 4$          & $5 \pm 4$          & $5.8 \pm 1.8$          & $2.4 \pm 1.1$          & $0.50 \pm 0.30$              \\

        Fitted $Z$+jets events         & $5.8 \pm 1.7$          & $1.8 \pm  0.6$ & $2.3 \pm 0.8$          & $1.00 \pm 0.30$          & $0.60 \pm 0.30$              \\

        Fitted single-top events         & $1.1 \pm 0.4$          & $0.10 \pm   0.10$          & $0.70 \pm 0.30$          & $0.20 \pm 0.10$          & $0.10 \pm 0.10$              \\

        Fitted diboson events         & $10 \pm 8$          & $2.5 \pm 1.7$          & $4 \pm 4$          & $2.3 \pm 1.5$          & $0.8 \pm 0.7$              \\

        Fitted \ttbar\+V events       & $0.34 \pm 0.08$          & $0.048 \pm 0.019$          & $0.20 \pm 0.10$          & $0.066 \pm 0.013$          & $0.025 \pm 0.007$              \\
\bottomrule
\end{tabular*}

\begin{tabular*}{\textwidth}{@{\extracolsep{\fill}}lcccc}

\toprule
\textbf{4J high-x} $b$-tag & All \meff\ bins & [1000,1500] & [1500,2000] &
$>2000$ [\GeV] \\
\midrule

Observed events          & $44$              & $38$              & $4$              & $2$                    \\
\midrule

Fitted bkg events         & $54 \pm 7$          & $44 \pm 6$          & $7.7 \pm 1.5$          & $1.8 \pm 0.7$              \\
\midrule

        Fitted \ttbar\ events         & $39 \pm 6$          & $34 \pm   6$          & $4.9 \pm 1.2$          & $0.8 \pm 0.5$              \\

        Fitted $W$+jets events         & $3.2 \pm 1.5$          & $2.7 \pm  1.5$ & $0.20 \pm 0.10$          & $0.20 \pm 0.10$              \\

        Fitted $Z$+jets events         & $0.14 \pm 0.13$          & $0.10  \pm 0.10$          & $0.036 \pm 0.025$          & $0.007_{-0.007}^{+0.019}$               \\

        Fitted single-top events         & $6.0 \pm 2.3$          & $4.1 \pm 2.2$          & $1.6 \pm 0.7$          & $0.30 \pm 0.30$              \\

        Fitted diboson events         & $1.6 \pm 1.0$          & $1.0 \pm 0.6$          & $0.30 \pm 0.30$          & $0.30  \pm 0.30$              \\

        Fitted \ttbar\+V events         & $3.69 \pm 0.25$          & $2.84 \pm 0.18$          & $0.70 \pm 0.10$          & $0.155 \pm 0.026$               \\

\toprule
\textbf{4J high-x} $b$-veto & All \meff\ bins & [1000,1500] & [1500,2000] &
$>2000$ [\GeV]\\
\midrule

Observed events          & $37$              & $27$              & $7$              & $3$                    \\
\midrule

Fitted bkg events         & $44 \pm 10$          & $36 \pm 10$          & $5.8 \pm 1.9$          & $2.4 \pm 1.0$              \\
\midrule

        Fitted \ttbar\ events         & $4.0 \pm 1.1$          & $3.3 \pm 1.0$    & $0.60 \pm 0.20$          & $0.10 \pm 0.10$              \\

        Fitted $W$+jets events         & $28 \pm 8$          & $24 \pm  8$          & $3.0 \pm 1.1$          & $1.0 \pm 0.4$              \\

        Fitted $Z$+jets events         & $1.4 \pm 0.8$          & $0.9 \pm  0.5$ & $0.10 \pm 0.10$          & $0.30 \pm 0.30$              \\

        Fitted single-top events         & $0.9 \pm 0.4$          & $0.6 \pm  0.4$          & $0.20 \pm 0.10$          & $0.10 \pm 0.10$              \\

        Fitted diboson events         & $9 \pm 5$          & $6.2 \pm 3.4$          & $1.8 \pm 1.5$          & $0.9 \pm 0.8$              \\

        Fitted \ttbar\+V events         & $0.36 \pm 0.09$          & $0.25 \pm 0.07$          & $0.057 \pm 0.018$          & $0.054 \pm 0.024$                \\

\bottomrule
\end{tabular*}

  \end{center}
\end{table}

\begin{table}
  \begin{center}
    \caption{\bgonlycaption{{\bf 4J low-x} and {\bf 6J} SRs}}

\begin{tabular*}{\textwidth}{@{\extracolsep{\fill}}lcccc}
\toprule
\textbf{4J low-x} $b$-tag & All \meff\ bins & [1300,1650] & [1650,2000] &
$>2000$ [\GeV] \\
\midrule

Observed events          & $31$              & $19$              & $6$              & $6$                    \\
\midrule

Fitted bkg events         & $31 \pm 6$          & $20 \pm 5$          & $6.6 \pm 2.3$          & $4.4 \pm 1.6$              \\
\midrule

        Fitted \ttbar\ events         & $19 \pm 5$          & $13 \pm        4$          & $4.3 \pm 1.6$          & $2.1 \pm 1.0$              \\

        Fitted $W$+jets events         & $2.4 \pm 1.1$          & $1.6 \pm        1.1$ & $0.40 \pm 0.20$          & $0.40 \pm 0.20$              \\

        Fitted $Z$+jets events         & $0.10 \pm 0.05$          & $0.037 \pm 0.033$          & $0.017 \pm 0.014$          & $0.042 \pm 0.030$               \\

        Fitted single-top events         & $6.5 \pm 3.4$          & $3.5 \pm        2.8$          & $1.5  \pm 1.5$          & $1.5 \pm 1.1$              \\

        Fitted diboson events         & $1.2 \pm 0.9$          & $0.7 \pm 0.4$          & $0.20 \pm 0.20$          & $0.30 \pm 0.30$              \\

        Fitted \ttbar\+V events         & $1.41 \pm 0.14$          & $1.03 \pm 0.12$          & $0.24 \pm 0.05$          & $0.14 \pm 0.04$              \\

\toprule
\textbf{4J low-x} $b$-veto & All \meff\ bins & [1300,1650] & [1650,2000] &
$>2000$ [\GeV] \\
\midrule

Observed events          & $19$              & $7$              & $7$              & $5$                    \\
\midrule

Fitted bkg events         & $16 \pm 4$          & $10.0 \pm 3.0$          & $3.5 \pm 1.1$          & $2.7 \pm 1.1$              \\
\midrule

        Fitted \ttbar\ events         & $2.3 \pm 0.8$          & $1.6 \pm 0.6$         & $0.40 \pm 0.20$          & $0.30 \pm 0.20$              \\

        Fitted $W$+jets events         & $7.5 \pm 2.6$          & $4.6 \pm        2.3$ & $1.9 \pm 0.8$          & $1.0 \pm 0.5$              \\

        Fitted $Z$+jets events         & $0.18 \pm 0.09$          &        $0.06_{-0.06}^{+0.06}$          & $0.05 \pm 0.04$          & $0.070 \pm 0.032$               \\

        Fitted single-top events         & $1.2 \pm 0.7$          & $0.5 \pm        0.4$          & $0.20 \pm 0.20$          & $0.5 \pm 0.4$              \\

        Fitted diboson events         & $4.9 \pm 3.3$          & $3.2 \pm 1.9$          & $1.0 \pm 0.7$          & $0.7_{-0.7}^{+0.9}$              \\

        Fitted \ttbar\+V events         & $0.125 \pm 0.031$          & $0.099 \pm 0.023$          & $0.014 \pm 0.008$          & $0.012_{-0.012}^{+0.014}$                \\
\bottomrule
\end{tabular*}

\begin{tabular*}{\textwidth}{@{\extracolsep{\fill}}lccccc}
\toprule
\textbf{6J} $b$-tag        & All \meff\ bins & [700,1233] & [1233,1767] &
[1767,2300] & $>2300$ [\GeV] \\
\midrule
Observed events          & $31$              & $6$              & $16$              & $9$              & $0$                    \\
\midrule

Fitted bkg events         & $27\pm 4$          & $8.5 \pm 2.0$          & $12.1 \pm 2.7$          & $4.1 \pm 1.0$          & $2.0 \pm 0.5$              \\
\midrule

        Fitted \ttbar\ events         & $19 \pm 4$          & $6.1 \pm        1.8$ & $9.1 \pm 2.5$          & $2.6 \pm 0.8$          & $1.1 \pm 0.4$              \\

        Fitted $W$+jets events         & $0.75 \pm 0.29$          & $0.20 \pm    0.20$ & $0.30 \pm 0.10$          & $0.20 \pm 0.10$          & $0.10 \pm 0.10$              \\

        Fitted $Z$+jets events         & $0.07 \pm 0.06$          & ... & $0.025_{-0.025}^{+0.026}$          & $0.029_{-0.029}^{+0.035}$          & $0.014 \pm 0.009$              \\

        Fitted single-top events         & $3.1 \pm 1.1$          & $0.7 \pm        0.5$          & $1.3 \pm 0.8$          & $0.6 \pm 0.4$          & $0.40 \pm 0.30$              \\

        Fitted diboson events         & $0.8 \pm 0.7$          & $0.20 \pm 0.20$          & $0.30 \pm 0.20$          & $0.20  \pm 0.20$          & $0.10  \pm 0.10$              \\

        Fitted \ttbar\+V events         & $2.9 \pm 0.5$          & $1.20 \pm   0.20$ & $1.10 \pm 0.30$          & $0.50 \pm 0.10$          & $0.2 \pm 0.0$              \\
\toprule
\textbf{6J} $b$-veto        & All \meff\ bins & [700,1233] & [1233,1767] &
[1767,2300] & $>2300$ [\GeV] \\
\midrule

Observed events          & $6$              & $3$              & $2$              & $1$              & $0$                    \\
\midrule

Fitted bkg events         & $7.3 \pm 2.0$          & $2.1 \pm 1.1$          & $3.4 \pm 0.8$          & $1.2 \pm 0.5$          & $0.60 \pm 0.30$              \\
\midrule

        Fitted \ttbar\ events         & $1.6 \pm 0.5$          & $0.50 \pm 0.20$         & $0.80 \pm 0.30$          & $0.20 \pm 0.10$          & $0.069 \pm 0.031$              \\

        Fitted $W$+jets events         & $2.4 \pm 0.9$          &        $0.6_{-0.6}^{+0.8}$          & $1.1 \pm 0.4$          & $0.50 \pm 0.30$          & $0.20 \pm 0.10$              \\

        Fitted $Z$+jets events         & $0.29 \pm 0.13$          & $0.1 \pm        0.0$ & $0.20 \pm 0.10$          & $0.026 \pm 0.015$          & $0.038 \pm 0.032$              \\

        Fitted single-top events         & $0.48 \pm 0.21$          & $0.10 \pm        0.10$          & $0.20 \pm 0.10$          & $0.10 \pm 0.10$          & $0.10 \pm 0.10$              \\

        Fitted diboson events         & $2.4 \pm 1.9$          & $0.7 \pm 0.7$          & $1.1 \pm 0.7$          & $0.4  \pm 0.4$          & $0.20  \pm 0.20$              \\

        Fitted \ttbar\+V events         & $0.13 \pm 0.04$          & $0.023 \pm 0.021$          & $0.089 \pm 0.021$          & $0.015 \pm 0.010$          & $0.0059 \pm 0.0021$              \\

\bottomrule
\end{tabular*}
  \end{center}
\end{table}

\begin{table}
  \begin{center}
    \caption{\label{results:tab:yieldsSRs_9J}\bgonlycaption{{\bf 9J} SR}}

\begin{tabular*}
{\textwidth}{@{\extracolsep{\fill}}lccc}
\toprule
\textbf{9J}                   & All \meff\ bins          & [1000,1500]  &
$>1500$ [\GeV] \\
\midrule
Observed events            & $10$                     & $6$                      & $4$                      \\
\midrule
Fitted bkg events          & $7\pm 4$           & $4.0\pm 2.6$           & $3.1\pm 1.6$           \\
\midrule
Fitted $W$+jets events     & ${0.028}_{-0.028}^{+0.057}$ & ${0.006}_{-0.006}^{+0.018}$ & ${0.022}_{-0.022}^{+0.043}$ \\
Fitted $t\bar t$ events    & $6\pm 4$           & $3.6\pm 2.5$           & $2.3\pm 1.6$           \\
Fitted single-top events    & $0.7\pm 0.6$      & ${0.18}_{-0.18}^{+0.27}$ & $0.5\pm 0.4$           \\
Fitted $t\bar t+V$ events  & $0.30\pm 0.17$           & $0.14\pm 0.09$           & $0.16\pm 0.09$           \\
Fitted diboson events      & ${0.15}_{-0.15}^{+0.16}$ &${0.029}_{-0.029}^{+0.041}$ & $ 0.12 \pm 0.12 $ \\
Fitted $Z$+jets events     & ${0.008}_{-0.008}^{+0.011}$ & ... & ${0.008}_{-0.008}^{+0.011}$ \\
\bottomrule
\end{tabular*}
  \end{center}
\end{table}

The top and $W$\,+\,jets background normalization factors obtained for the \textbf{2J}, \textbf{4J low-x}, \textbf{4J high-x}, and \textbf{6J} SRs are shown in bins of \meff\ in Figure~\ref{results:NF}.
A trend toward smaller normalization factors at large values of \meff\ is
observed, which demonstrates the necessity of applying the same binning
requirements in control and signal regions.
The predicted event yields from \ttbar\ events in which both top quarks decay semileptonically
are cross-checked using the alternative object-replacement method described in Section~\ref{sec:background}. 
Figure~\ref{results:objectReplacementPulls} shows that the background estimates
obtained from the two methods are consistent.
Figures~\ref{results:meffbv}--\ref{results:meffbt} show the \meff\ distribution in {\bf 2J}, {\bf 4J low-x}, {\bf 4J high-x} and {\bf 6J} in $b$-tag and $b$-veto signal regions after fit. Figure~\ref{results:meff9J} shows the \meff\ distribution in {\bf 9J} signal region after fit.
The uncertainty bands plotted include all statistical and systematic uncertainties. The dashed lines stand for the benchmark signal samples.

Using the results of the background-only fit, a \textit{model-independent limit}
fit is performed to test for the presence of any beyond-the-Standard-Model (BSM)
physics processes that contribute to the SR (``disc'' SR in
Table~\ref{tab:SR1}) .
The BSM signal is assumed to contribute only to the SR and not to the CRs, thus giving
a conservative estimate of background in the SR. 
Observed ($S_\mathrm{obs}^{95}$) and expected ($S_\mathrm{exp}^{95}$)
95\% confidence level (CL) upper limits on the number of BSM signal events
are derived using the CL$_{\rm s}$ prescription~\cite{Read:2002hq}.
Table~\ref{results:tab:upperlimits} presents these limits,
together with the upper limits on the visible BSM cross-section,
$\langle\epsilon\sigma\rangle_{\mathrm{obs}}^{95}$, defined as the product
of acceptance, selection efficiency and production cross-section. 
The upper limits on the visible BSM cross-section are calculated by dividing the observed upper limit on
the beyond-SM events by the integrated luminosity of $36.1$ \ifb.
Moreover, the discovery $p$-values are given. They quantify the probability under the background-only 
hypothesis to produce event yields greater than or equal to the observed data.

Additionally, the results are interpreted in the specific supersymmetric scenarios
described in Section~\ref{sec:samples} using \textit{model-dependent limit} fits.
A \textit{model-dependent limit} fit takes the data event yields in multiple, 
statistically independent SRs 
and their associated CRs to compute an upper limit on the cross-section of a
targeted SUSY model.
The fit includes the expected signal contributions to the SRs and to the CRs,
scaled by a floating signal normalization factor.
The background normalization factors are also determined simultaneously in the fit.

The sparticle mass in a specific SUSY model can be excluded if the upper limit
of the signal normalization factor obtained in the fit is smaller than unity.

For the gluino/squark one-step models,
a model-dependent fit is performed over all bins of the \textbf{2J}, \textbf{4J high-x}, \textbf{4J low-x}, and \textbf{6J} SRs.
An independent set of background normalization factors are allocated 
for each bin of each SR (``excl'' SR in Table~\ref{tab:SR1}) and its associated CRs.
Figure~\ref{sec:results:contours} (top and middle) shows the observed
and expected exclusion bounds at 95\% CL for the one-step simplified models with gluino and
squark production. Gluino masses up to 2.1~\TeV~and
squark masses up to 1.25~\TeV~are excluded. 

Figure~\ref{sec:results:contours} (bottom) shows the exclusion contours of the
\textbf{9J} SR (Table~\ref{tab:SR2}) for the gluino two-step as well as the pMSSM scenario
described in Section~\ref{sec:samples}. In both cases the limits reach well
beyond 1.7~\TeV~in gluino mass.

\begin{figure}[h!]
  \begin{center}
    \includegraphics[width=\textwidth]{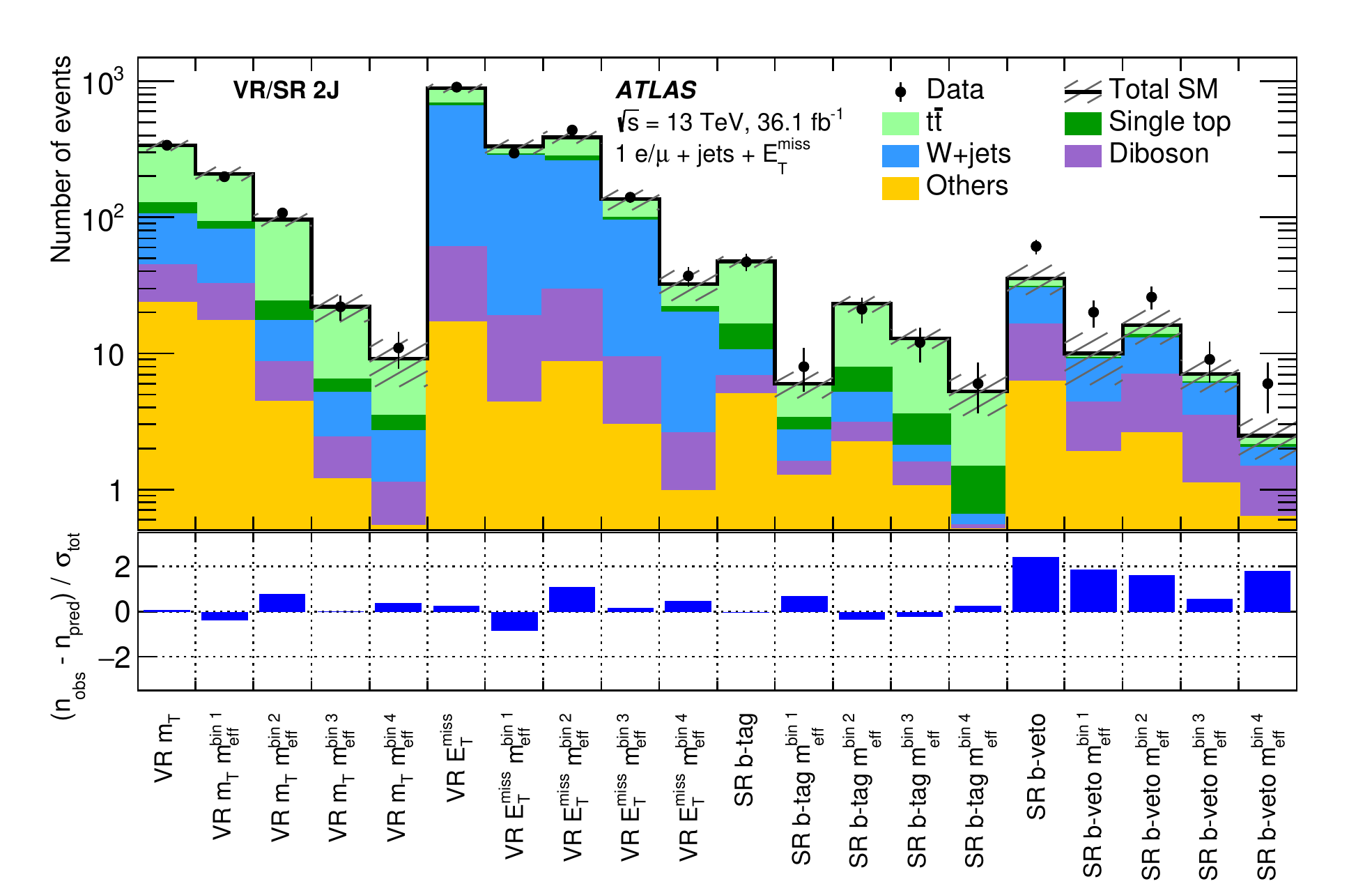}
  \end{center}
  \caption{Comparison of the observed and expected event yields in the
  \textbf{2J} validation and signal regions. Uncertainties in the background
  estimates include both the statistical (in the simulated event yields) and
  systematic uncertainties. Both the integrated regions and the regions for each \meff\ bin are presented.}
  \label{results:2JVRSRPulls}
\end{figure}

\begin{figure}[h!]
  \begin{center}
    \includegraphics[width=\textwidth]{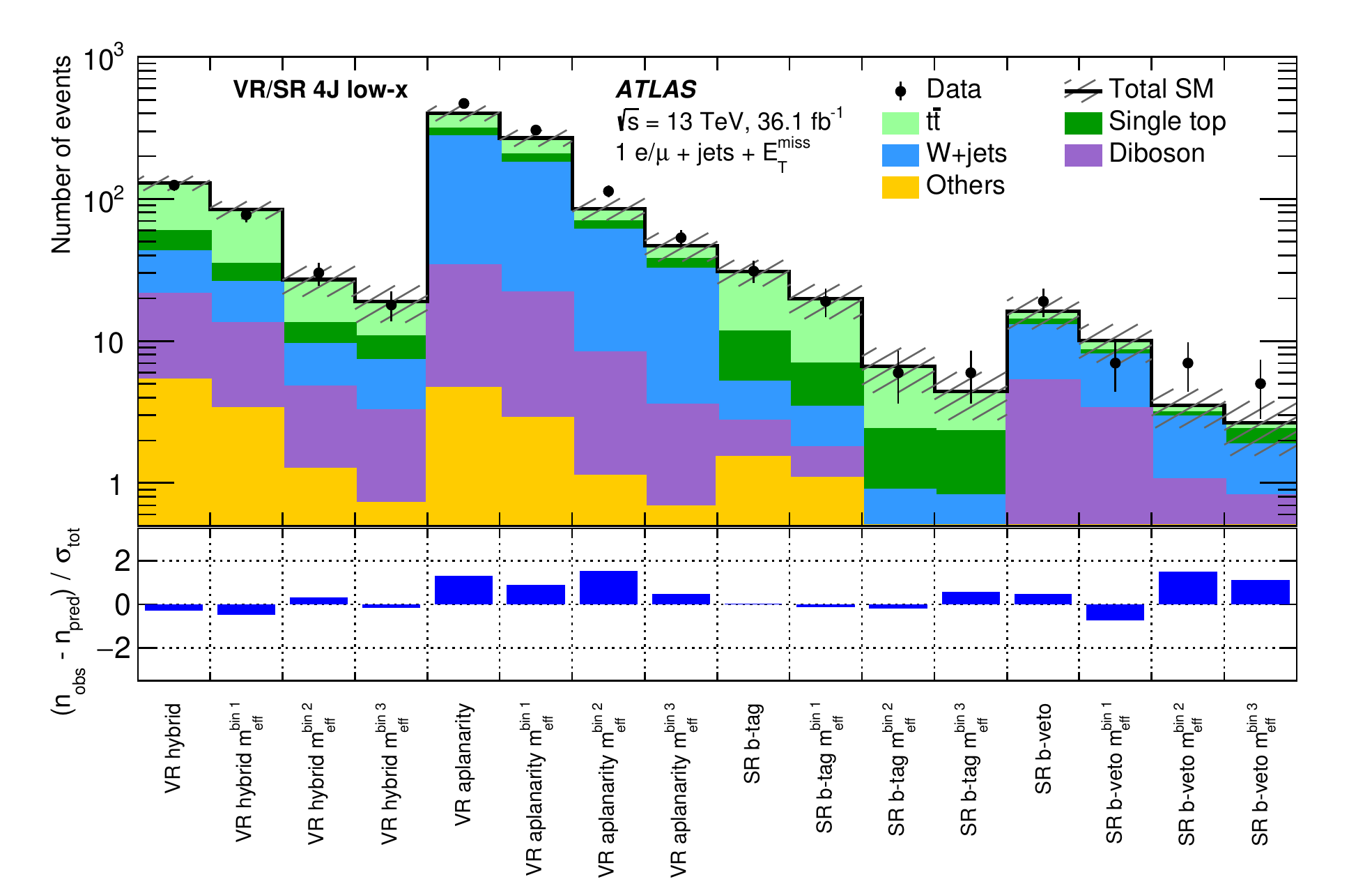}
  \end{center}

  \caption{Comparison of the observed and expected event yields in the
  \textbf{4J low-x} validation and signal regions. Uncertainties in the
  background estimates include both the statistical (in the simulated event
  yields) and systematic uncertainties. Both the integrated regions and the regions for each \meff\ bin are presented.}
  \label{results:4JlowxVRSRPulls}
\end{figure}

\begin{figure}[h!]
  \begin{center}
    \includegraphics[width=\textwidth]{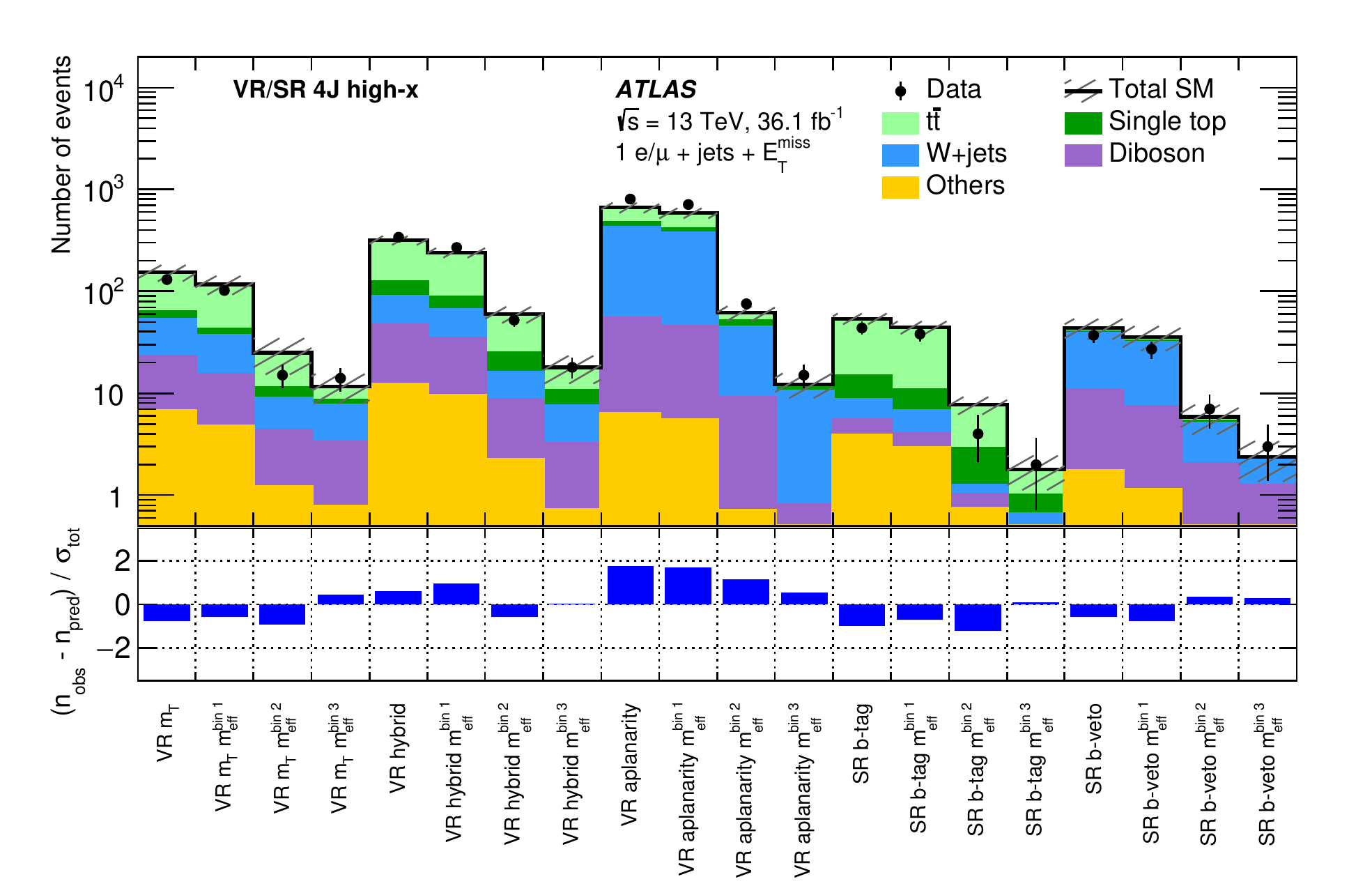}
  \end{center}

  \caption{Comparison of the observed and expected event yields in the
  \textbf{4J high-x} validation and signal regions. Uncertainties in the
  background estimates include both the statistical (in the simulated event
  yields) and systematic uncertainties. Both the integrated regions and the regions for each \meff\ bin are presented.}
  \label{results:4JhighxVRSRPulls}
\end{figure}

\begin{figure}[h!]
  \begin{center}
    \includegraphics[width=\textwidth]{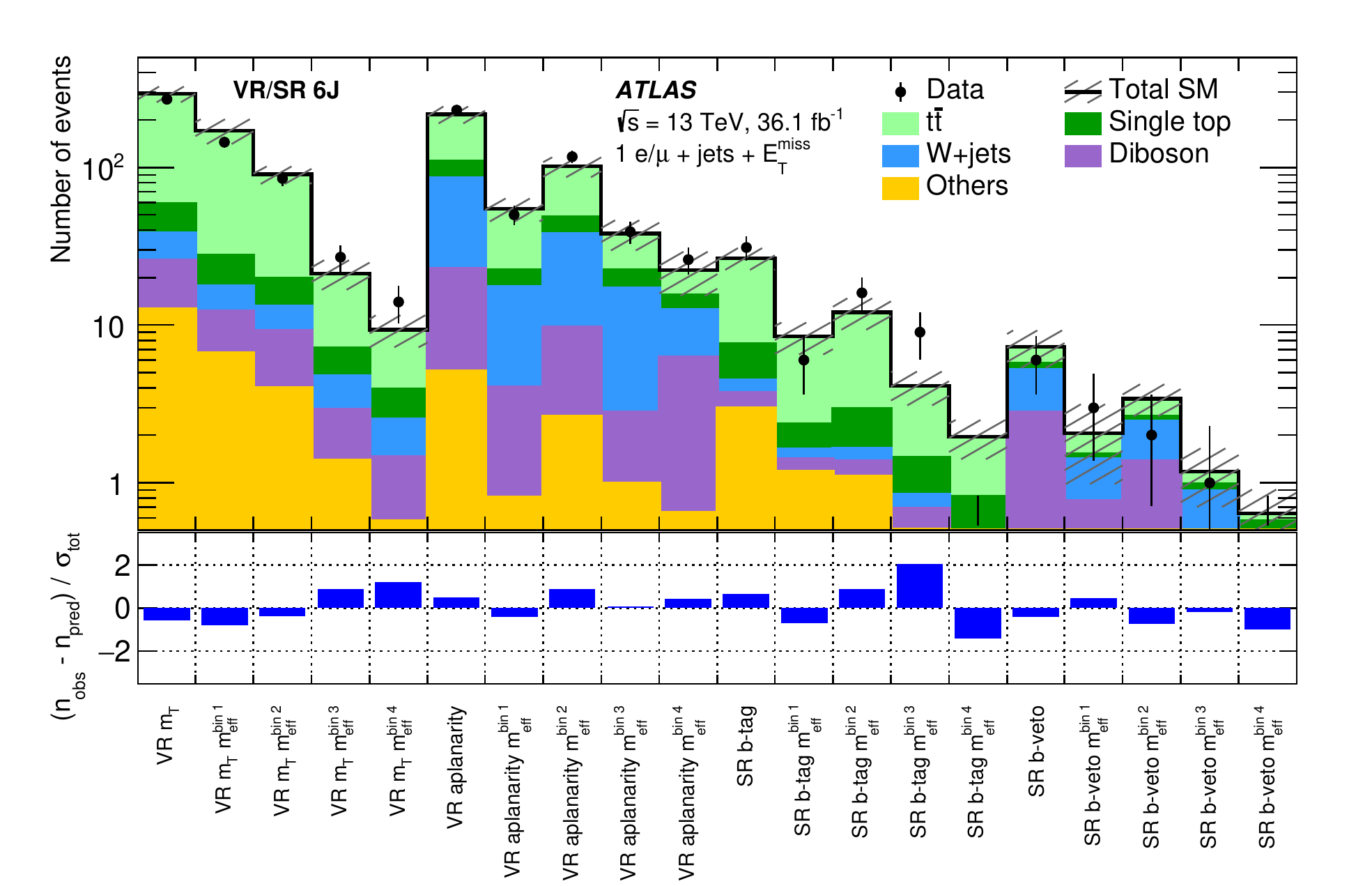}
  \end{center}

  \caption{Comparison of the observed and expected event yields in the
  \textbf{6J} validation and signal regions. Uncertainties in the background
  estimates include both the statistical (in the simulated event yields) and
  systematic uncertainties. Both the integrated regions and the regions for each \meff\ bin are presented.}
  \label{results:6JVRSRPulls}
\end{figure}

\begin{figure}[h!]
  \begin{center}
    \includegraphics[width=\textwidth]{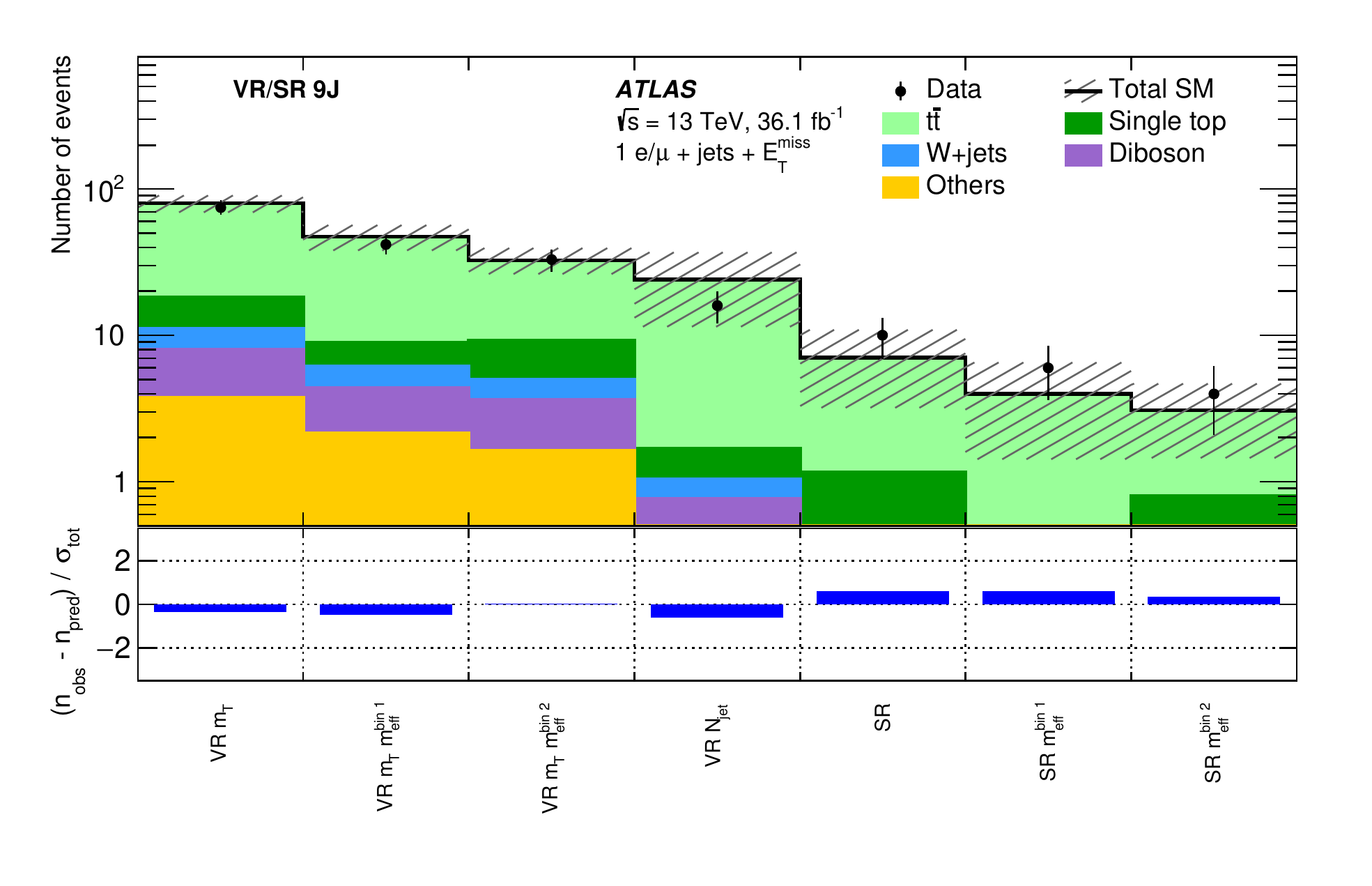}
  \end{center}

  \caption{Comparison of the observed and expected event yields in the
  \textbf{9J} validation and signal regions. Uncertainties in the background
  estimates include both the statistical (in the simulated event yields) and
  systematic uncertainties. Both the integrated regions and the regions for each \meff\ bin are presented.}
  \label{results:9JVRSRPulls}
\end{figure}

\begin{figure}[h!]
  \begin{center}
    \includegraphics[width=0.49\textwidth, angle =90 ]{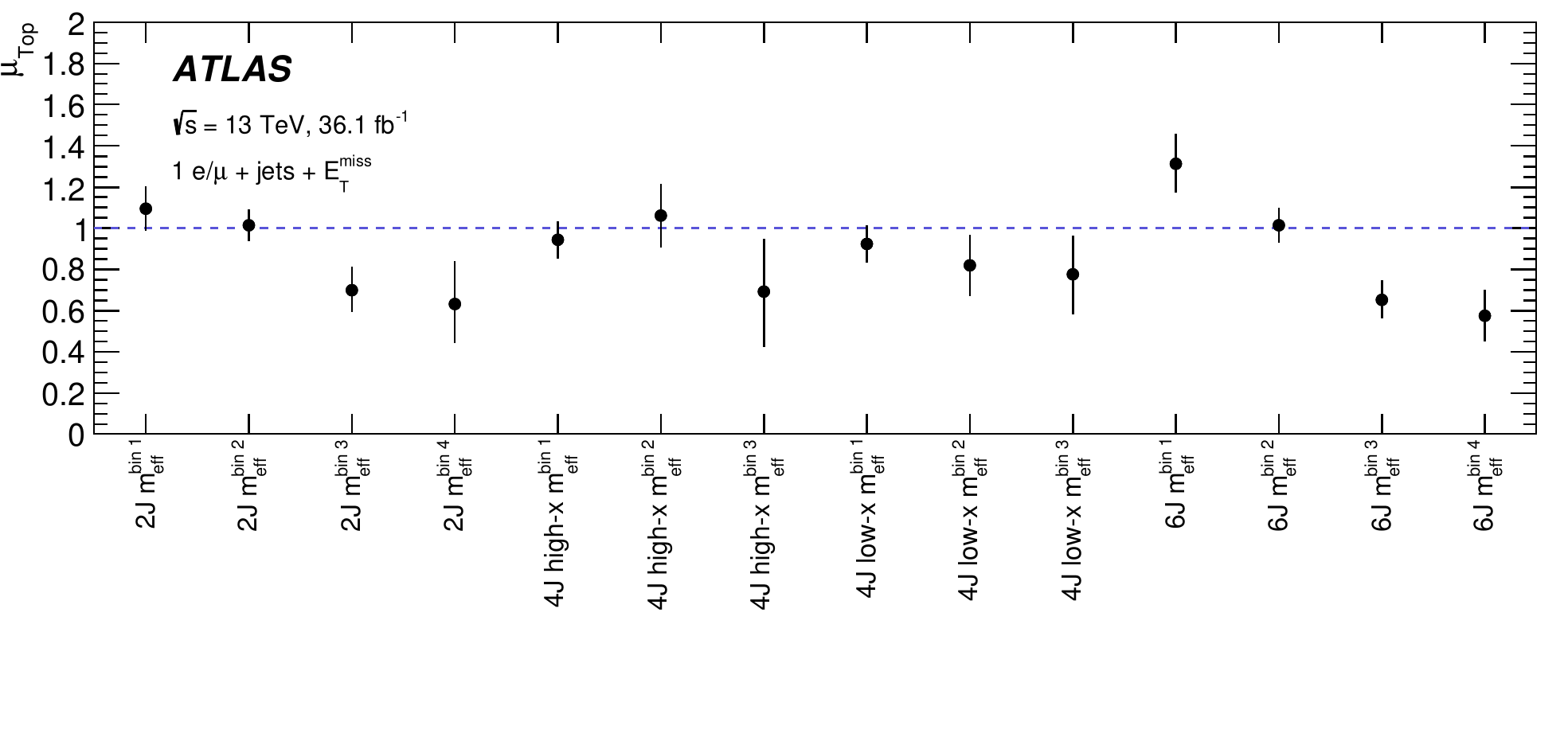}
    \includegraphics[width=0.49\textwidth, angle =90 ]{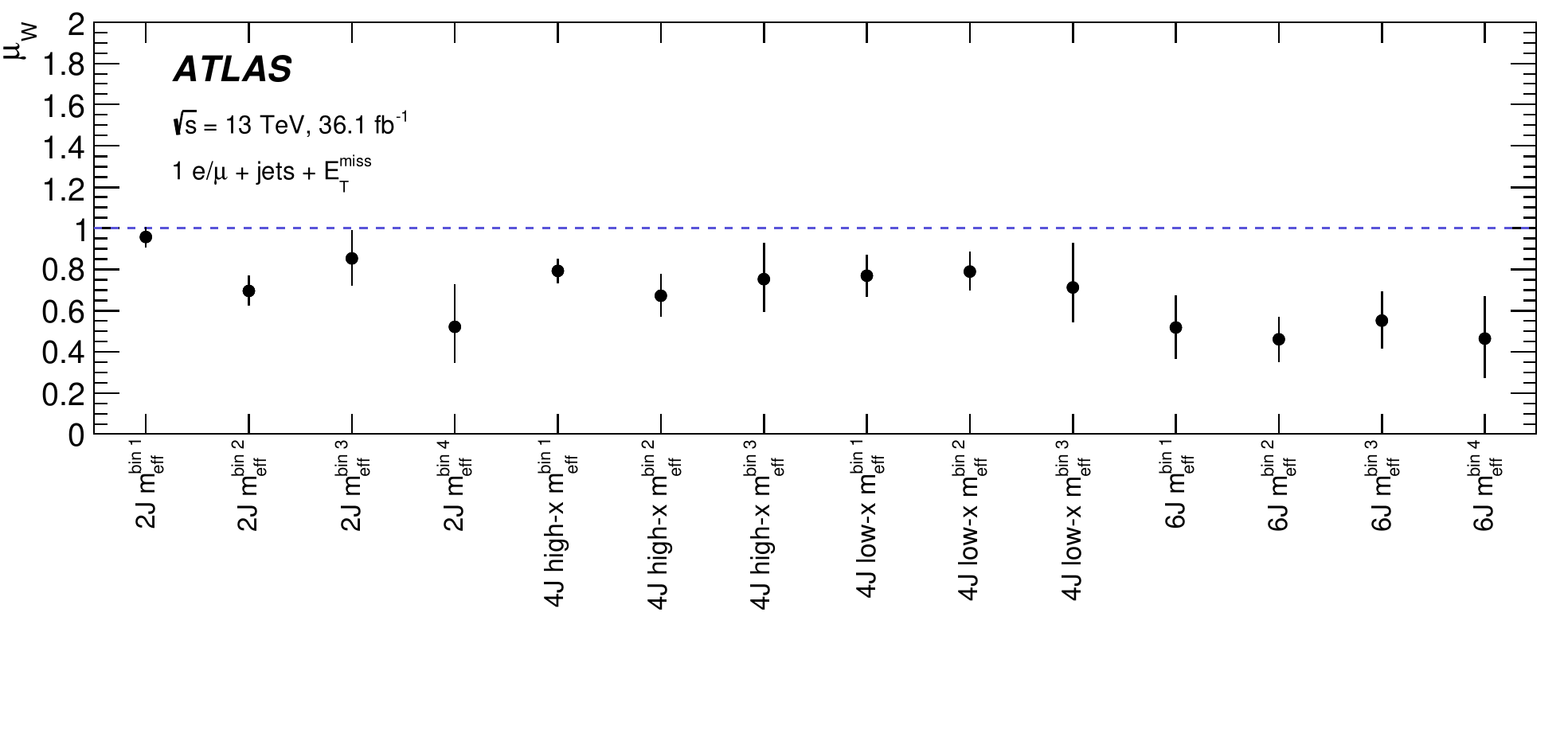}
  \end{center}
  \caption{\label{results:NF}
    Normalization factors $\mu_{\text{top}}$ and $\mu_{W}$ of the top and
    $W$\,+\,jets backgrounds in the \meff\ bins of the {\bf 2J}, {\bf 4J}
    \mbox{\bf high-x}, {\bf 4J low-x}, and {\bf 6J} SRs obtained with the
    background-only fit.
    The error bars correspond to the uncertainties as derived in the fit.}
\end{figure}

\begin{figure}[h!]
  \begin{center}
    \includegraphics[width=0.65\textwidth, angle =90 ]{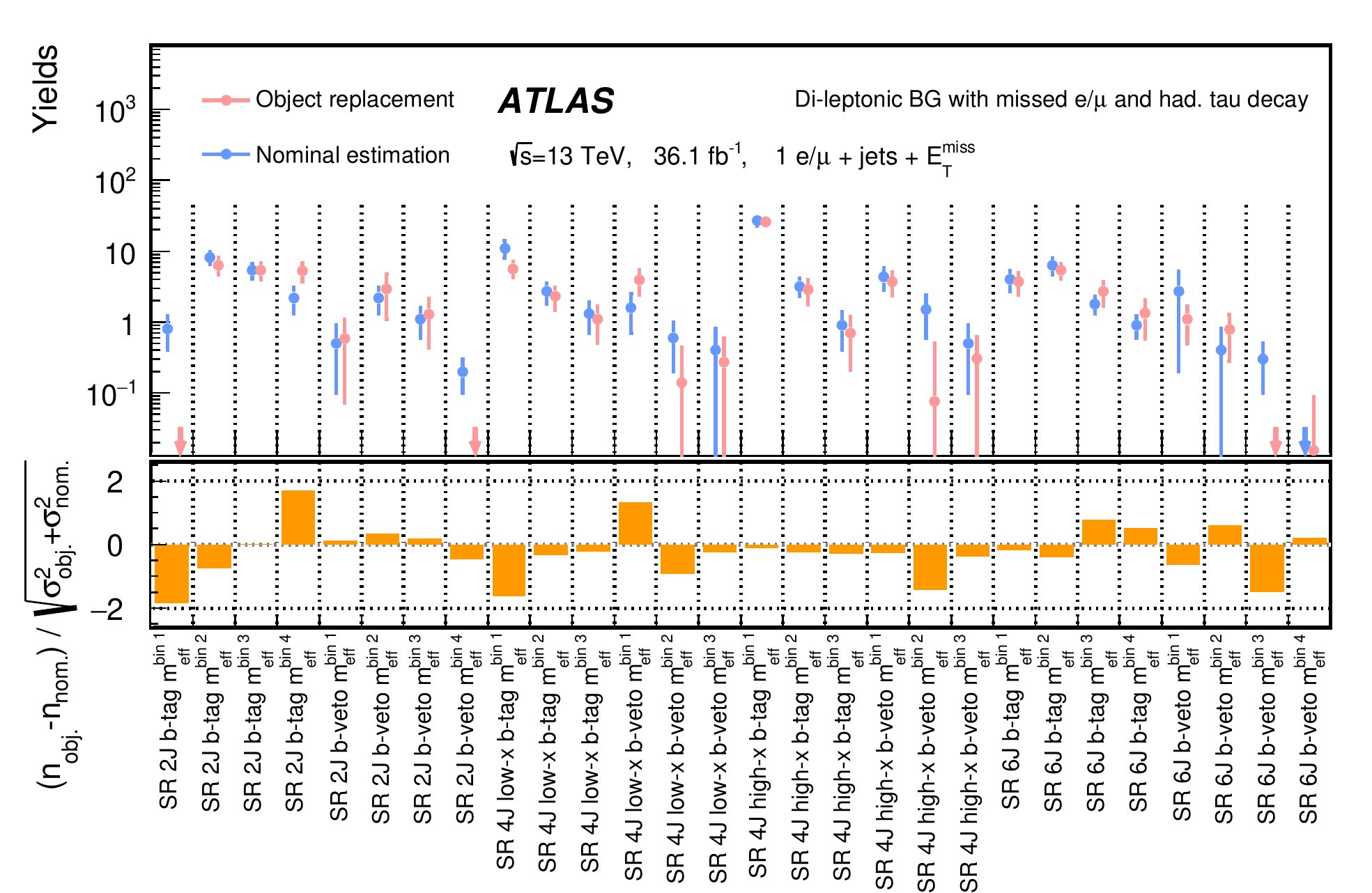}
  \end{center}
  \caption{\label{results:objectReplacementPulls}
	Comparison of estimated Standard Model background event yield from
	\ttbar\ processes with dileptonic decays in the {\bf 2J}, {\bf 4J high-x}, {\bf
	4J low-x}, and {\bf 6J} SR using the nominal semi-data-driven background
	estimation technique and the alternative object-replacement 
    technique described in Section~\ref{sec:background}. The error bands include
    both the statistical and systematic uncertainties.}
\end{figure}

\begin{table}
\caption{\label{results:tab:upperlimits}
	Results of the model-independent limit fits.
	For each SR, the observed 95\% CL upper limit on the visible cross-section 
	($\langle\epsilon\sigma\rangle_{\mathrm{obs}}^{95}$), 
        the observed ($S_{\mathrm{obs}}^{95}$) and expected ($S_{\mathrm{exp}}^{95}$) 95\% CL upper limits on the BSM event yield,
	and the one-sided discovery $p$-value ($p(s = 0)$) are presented.
	The $p$-values are capped at $0.5$ if fewer events than the 
	fitted background estimate are observed.}
\begin{center}
\begin{tabular*}{\textwidth}{@{\extracolsep{\fill}}l|ccccccc}
\toprule
\textbf{SR$_{\mathrm{disc}}$} &\textbf{2J}& \textbf{4J high-x}& \textbf{4J low-x}  & \textbf{4J low-x}  & \textbf{6J}   & \textbf{6J}    & \textbf{9J}      \\
             &  &          &          (gluino) &          (squark)&    (gluino)&    (squark) & \\
\midrule

Observed events           & $80$  &    $16$  &    $24$  &    $50$  &    $0$  &    $28$  &  $4$ \\
Fitted bkg events          & $67 \pm 6$ & $17.7 \pm 2.7$ & $17.2 \pm 3.2$ & $47 \pm 7$ & $2.6 \pm 0.6$ & $23.4 \pm 3.1$ & $3.1 \pm 1.6$  \\
$\langle\epsilon\sigma\rangle_{\mathrm{obs}}^{95}$ [fb]  &$0.92$ &$0.27$ &$0.50$ &$0.62$ &$0.08$ & $0.46$  & $0.20$ \\
$S_{\mathrm{obs}}^{95}$ &$33.1$ &$9.8$  &$18.0$ &$22.5$ &$3.0$  & $16.6$ & $7.1$\\
$S_{\mathrm{exp}}^{95}$ &$ { 21.6 }^{ +9.2 }_{ -5.6 }$  &$ { 10.8 }^{ +3.7 }_{ -3.0 }$  &$ { 11.8 }^{ +4.8 }_{ -2.7 }$  &$ { 19.9 }^{ +7.5 }_{ -5.6 }$  &$ { 4.5 }^{ +1.8 }_{ -1.0 }$   & $ { 12.7 }^{ +5.0 }_{ -4.0 }$ & ${ 6.0 }^{ +2.2 }_{ -1.2 }$ \\
$p(s=0)$ & $ 0.10$& $ 0.50$& $ 0.10$& $ 0.35$& $ 0.50$&$ 0.21$ & $ 0.34$\\

\bottomrule
\end{tabular*}
\end{center}
\end{table}

\begin{figure}[tbh]
  \begin{center}

    \includegraphics[width=0.49\textwidth]{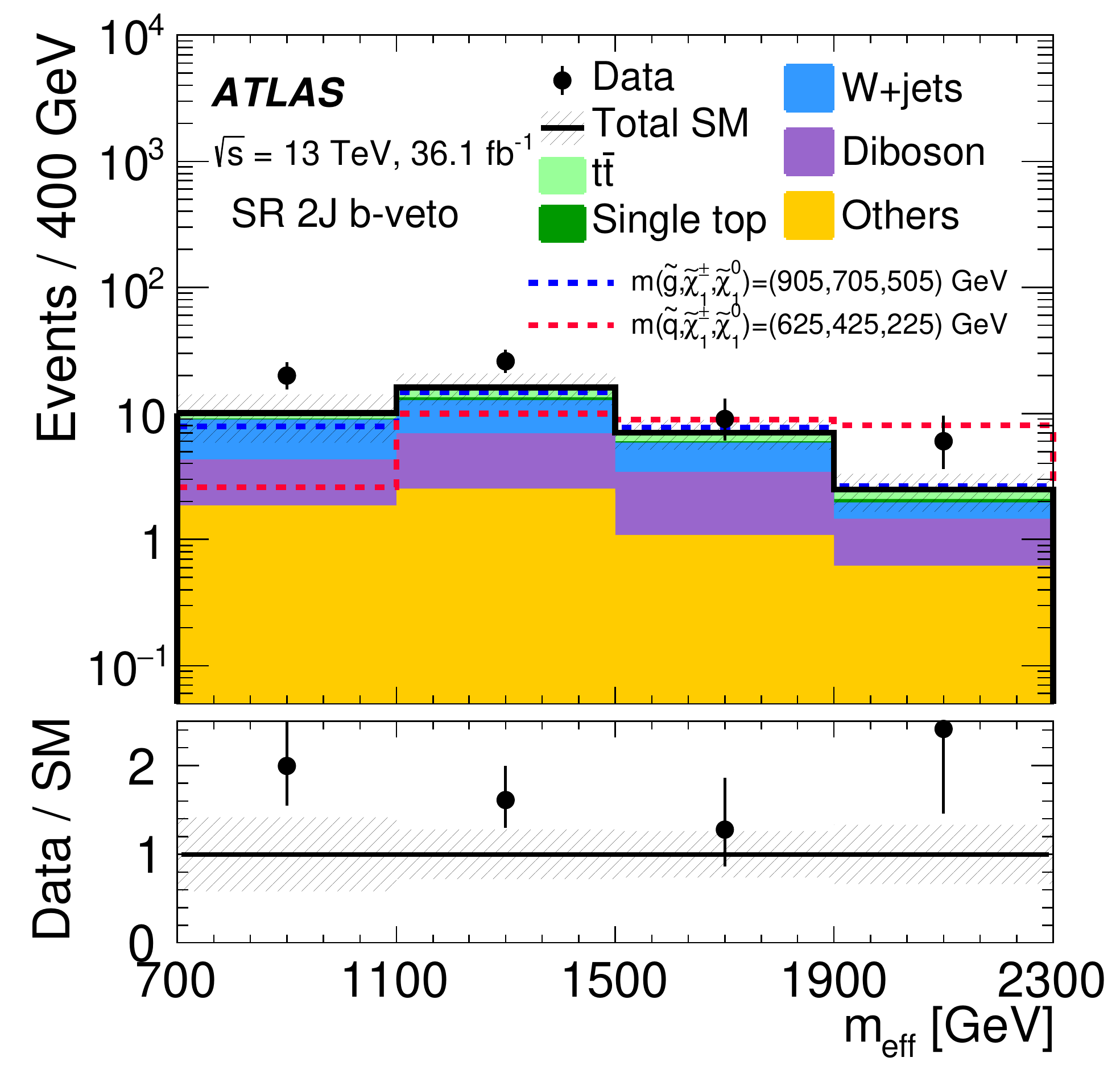}
    \includegraphics[width=0.49\textwidth]{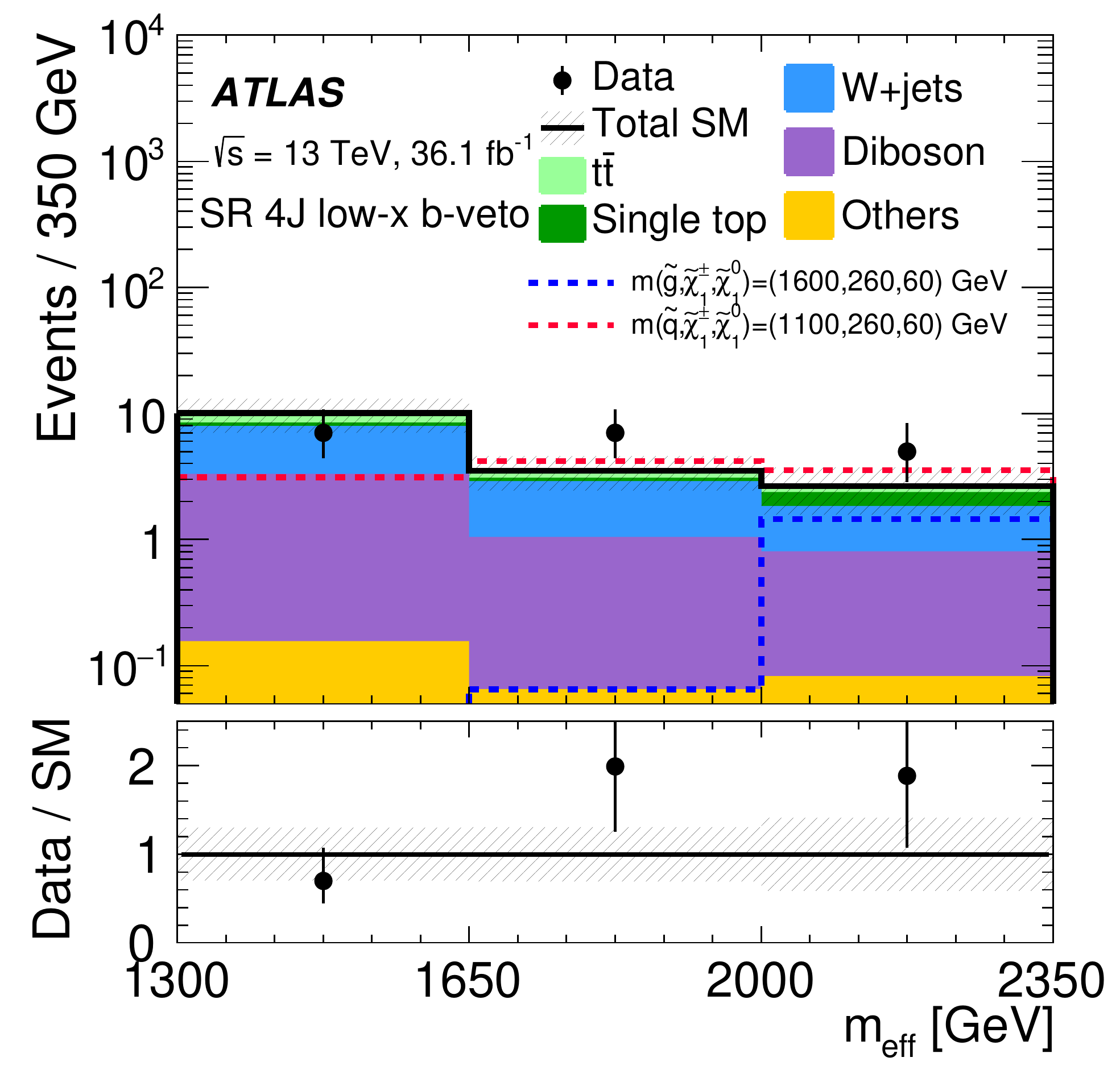}
    \includegraphics[width=0.49\textwidth]{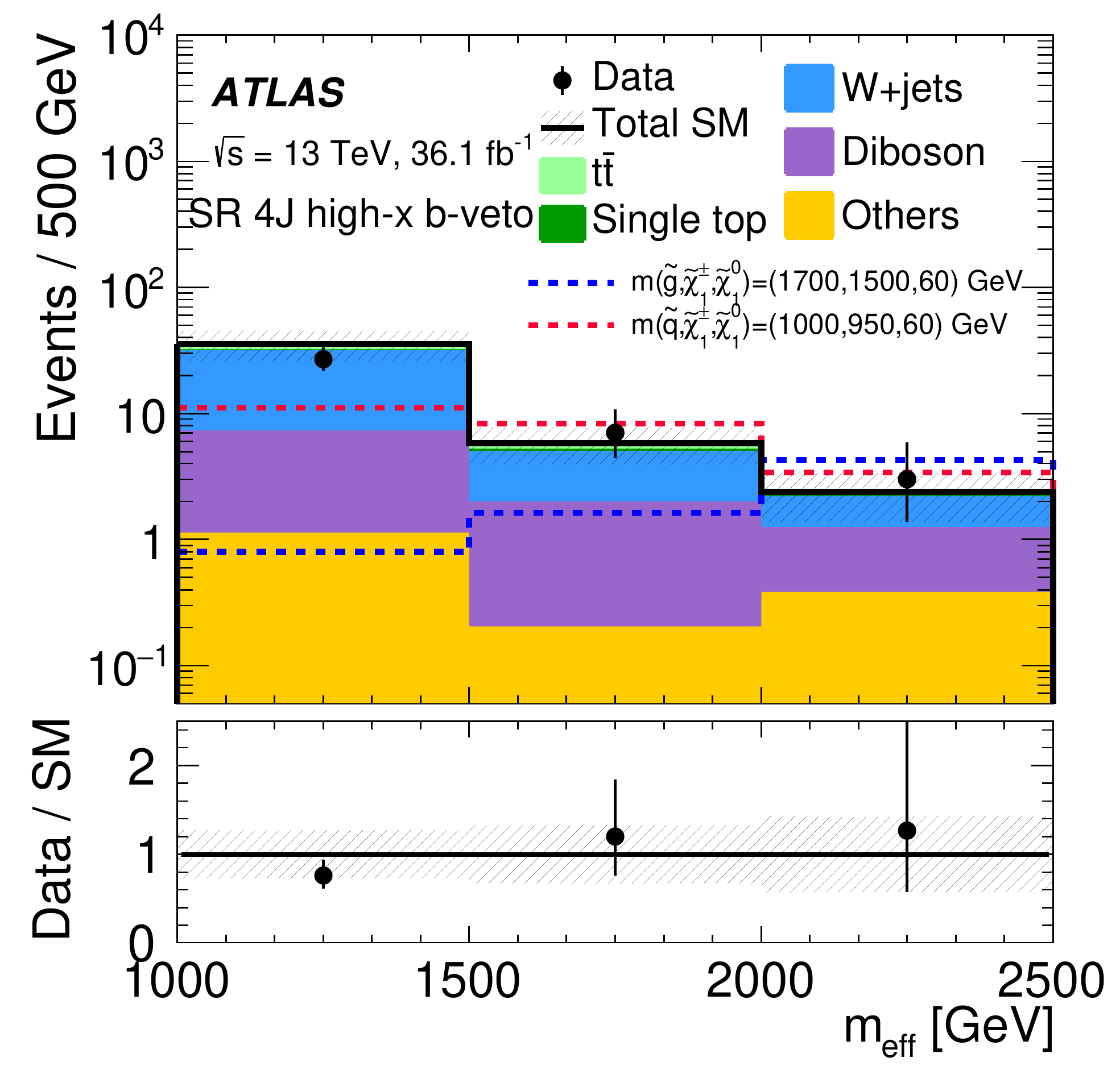}
    \includegraphics[width=0.49\textwidth]{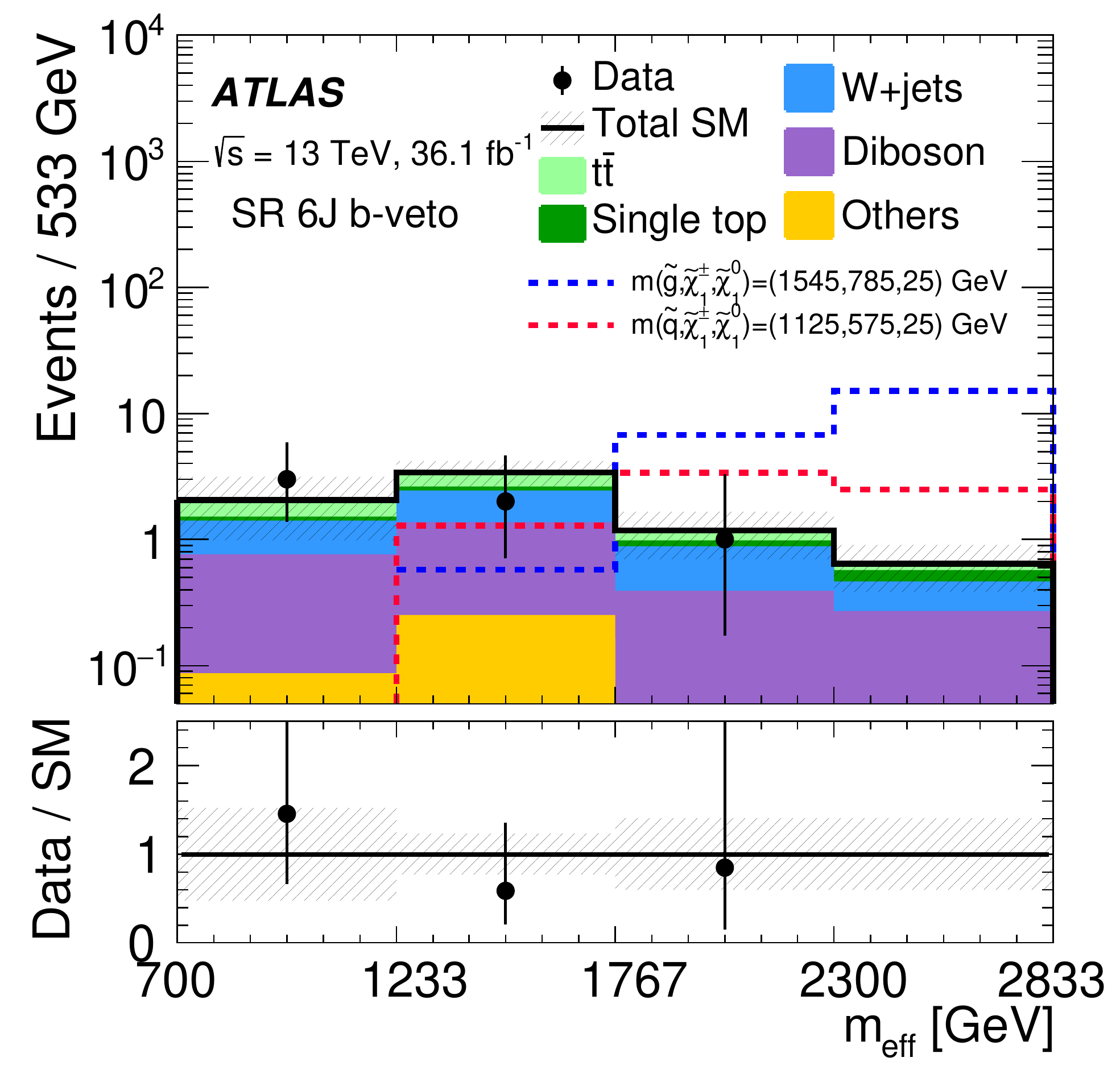}
  \end{center}
  \caption{\label{results:meffbv} The \meff\ distribution in {\bf 2J} (top left),
  {\bf 4J low-x} (top right), {\bf 4J high-x} (bottom left) and {\bf 6J} (bottom
  right) $b$-veto signal regions after fit. The uncertainty bands plotted include all statistical and systematic uncertainties. The dashed lines stand for the benchmark signal samples.}
\end{figure}

\begin{figure}[tbh]
  \begin{center}
    \includegraphics[width=0.49\textwidth]{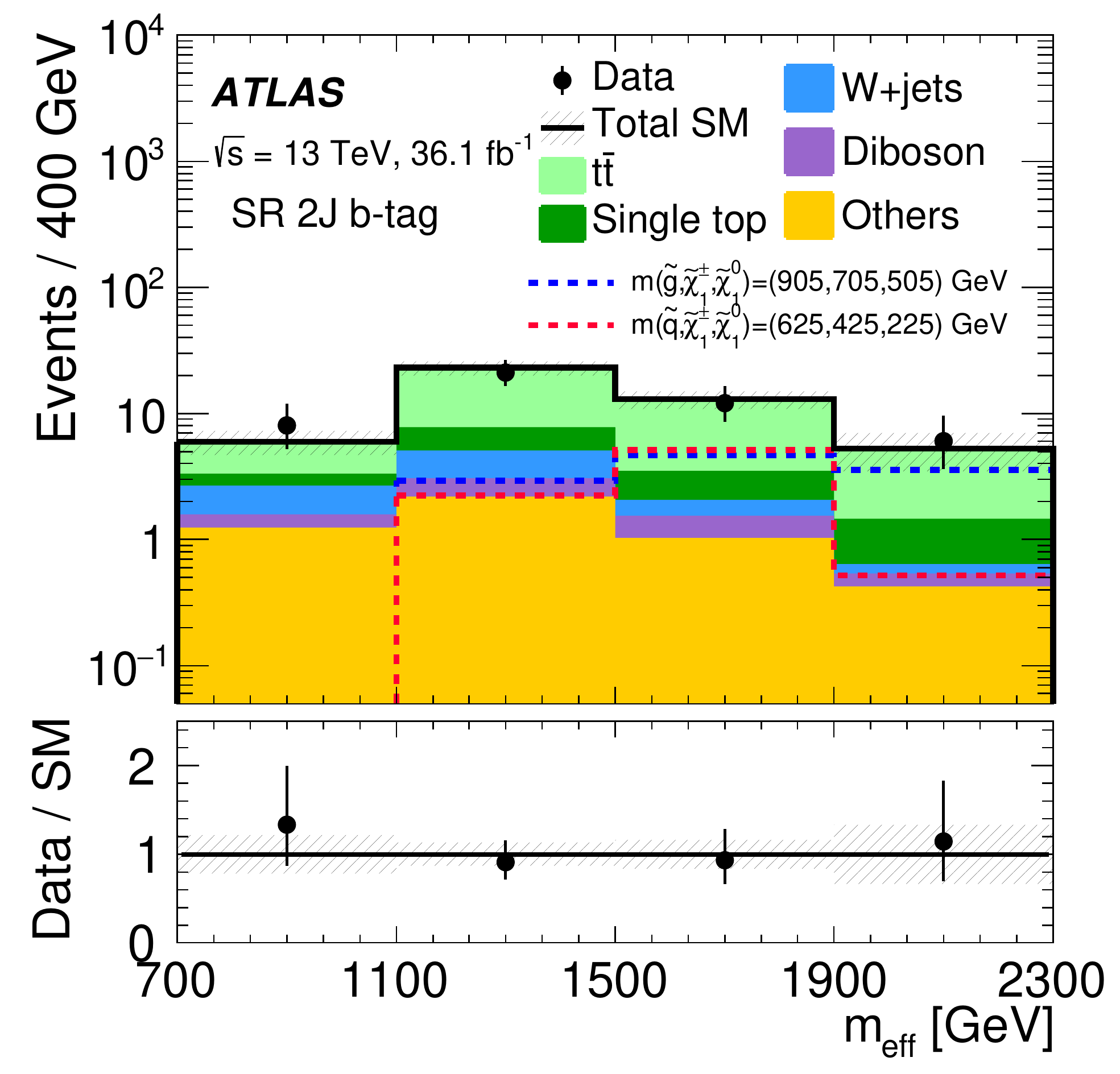}
    \includegraphics[width=0.49\textwidth]{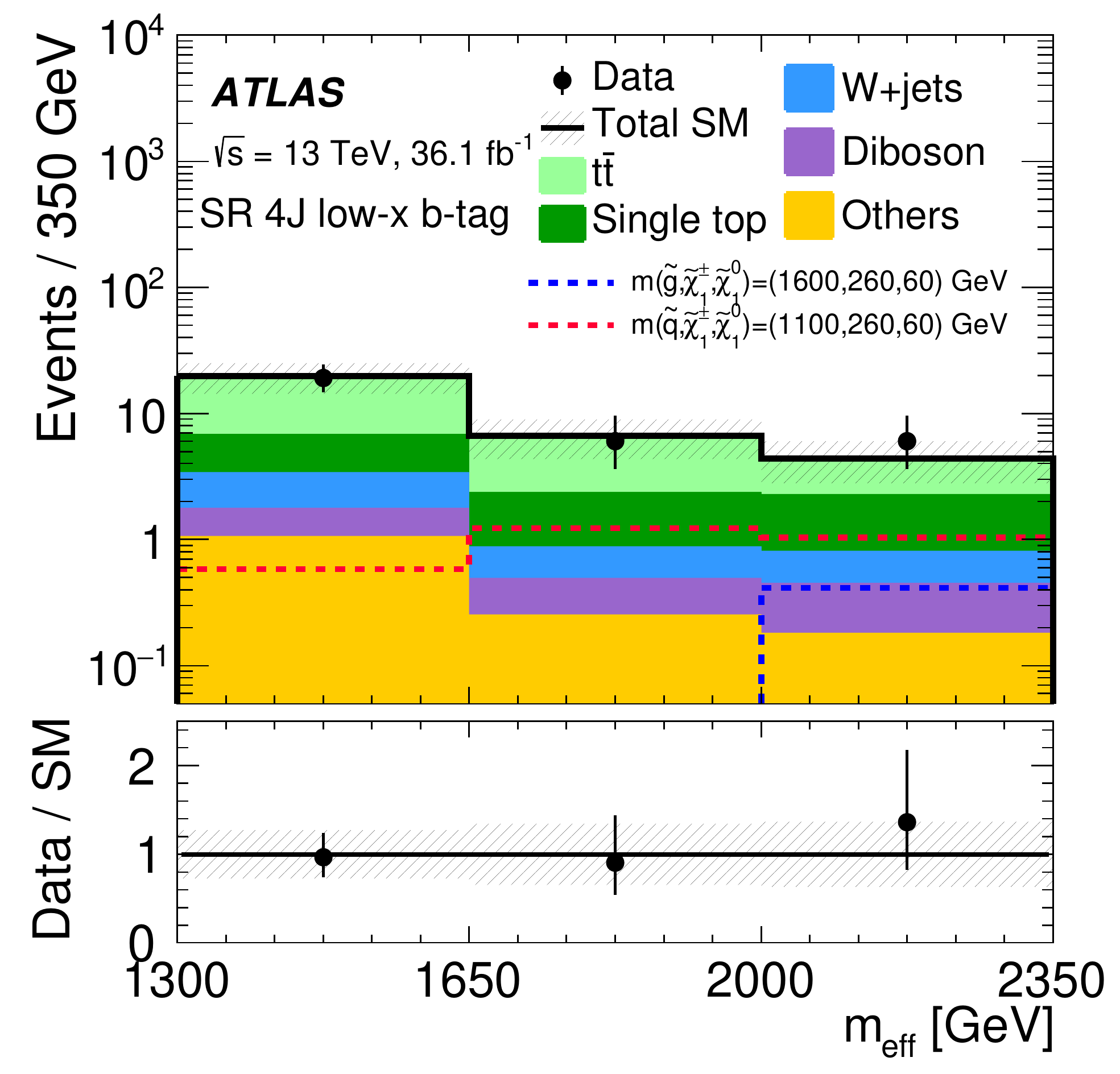}
    \includegraphics[width=0.49\textwidth]{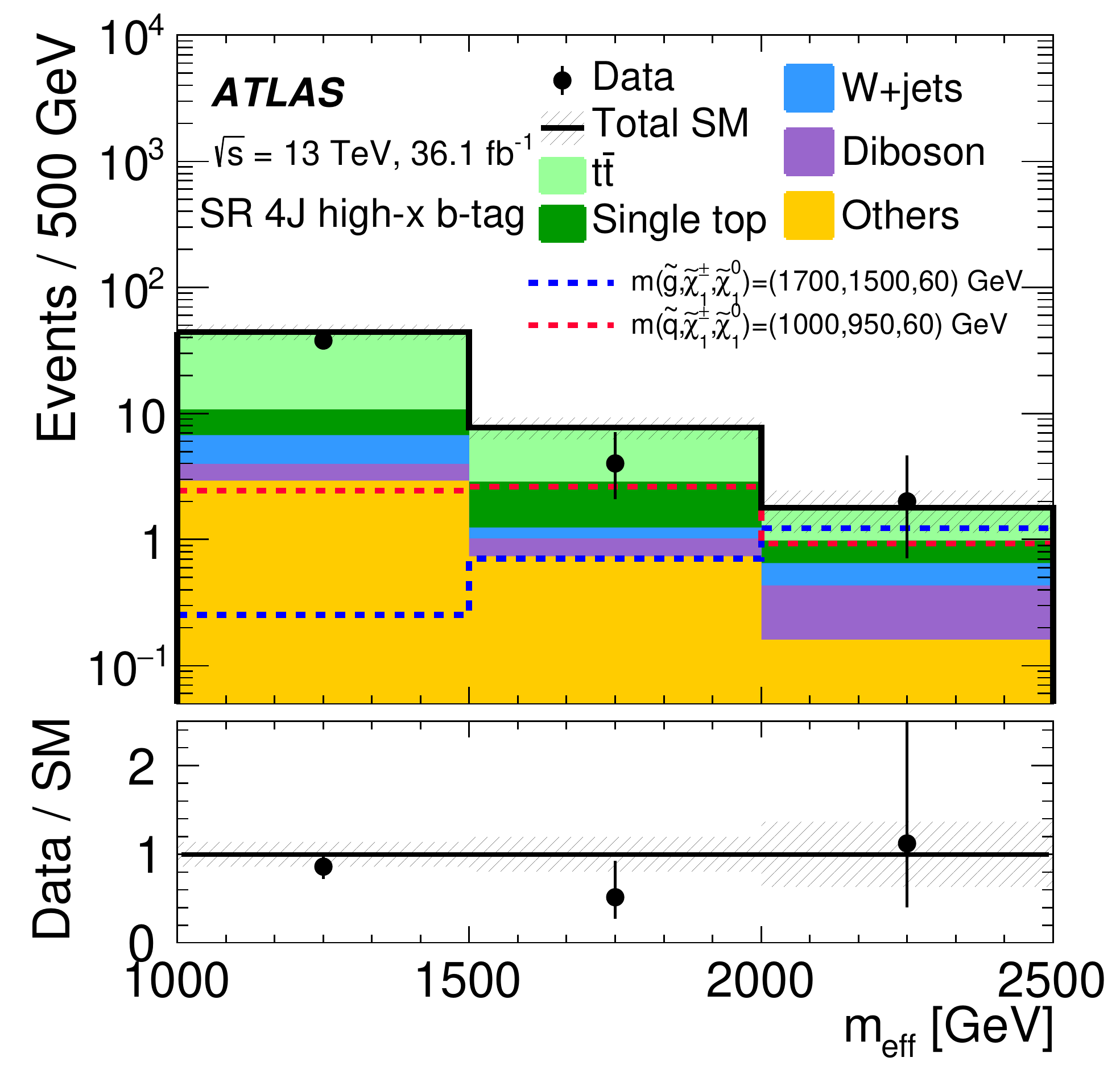}
    \includegraphics[width=0.49\textwidth]{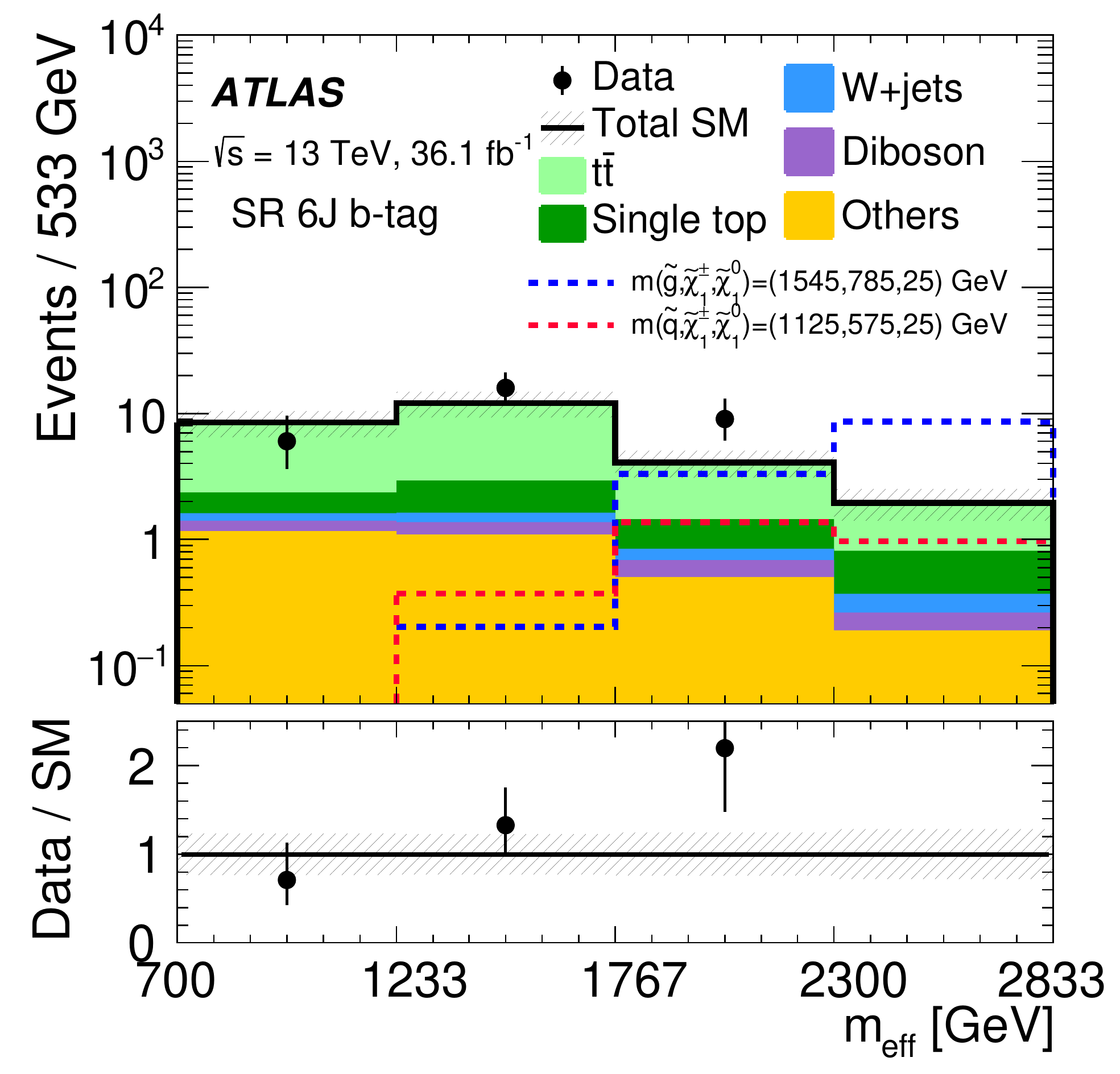}
  \end{center}
  \caption{\label{results:meffbt} The \meff\ distribution in {\bf 2J} (top left),
  {\bf 4J low-x} (top right), {\bf 4J high-x} (bottom left) and {\bf 6J} (bottom
  right) $b$-tag signal regions after fit.  The uncertainty bands plotted include all statistical and systematic uncertainties.  The dashed lines stand for the benchmark signal samples.}
\end{figure}

\begin{figure}[tbh]
  \begin{center}
    \includegraphics[width=0.49\textwidth]{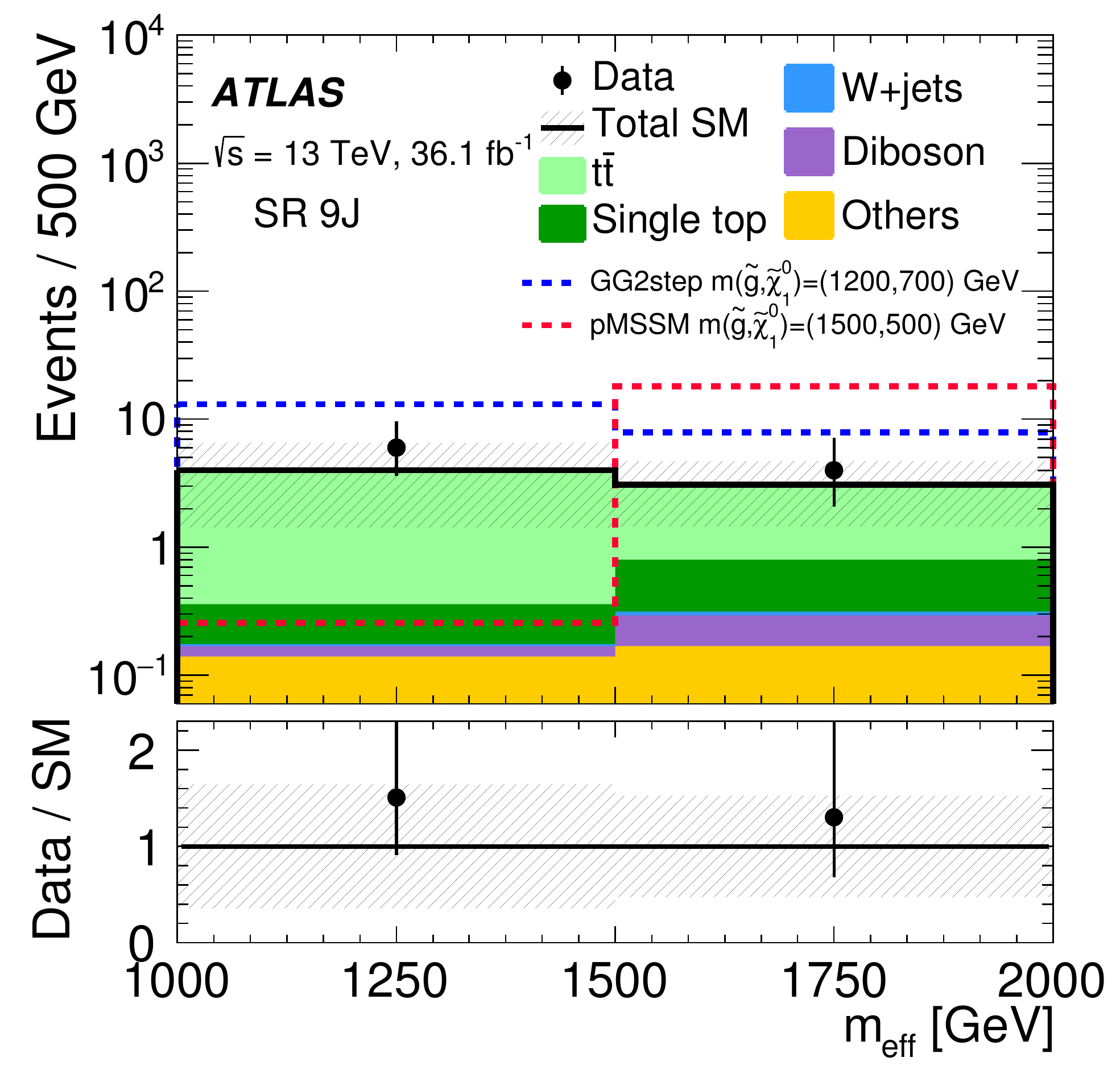}
  \end{center}
  \caption{\label{results:meff9J} The \meff\ distribution in the {\bf 9J} signal
  region  after fit. The uncertainty bands plotted include all statistical and systematic uncertainties. The dashed lines stand for the benchmark signal samples.}
\end{figure}

\begin{figure}[bh!]
 \begin{center}
  \includegraphics[width=0.49\textwidth]{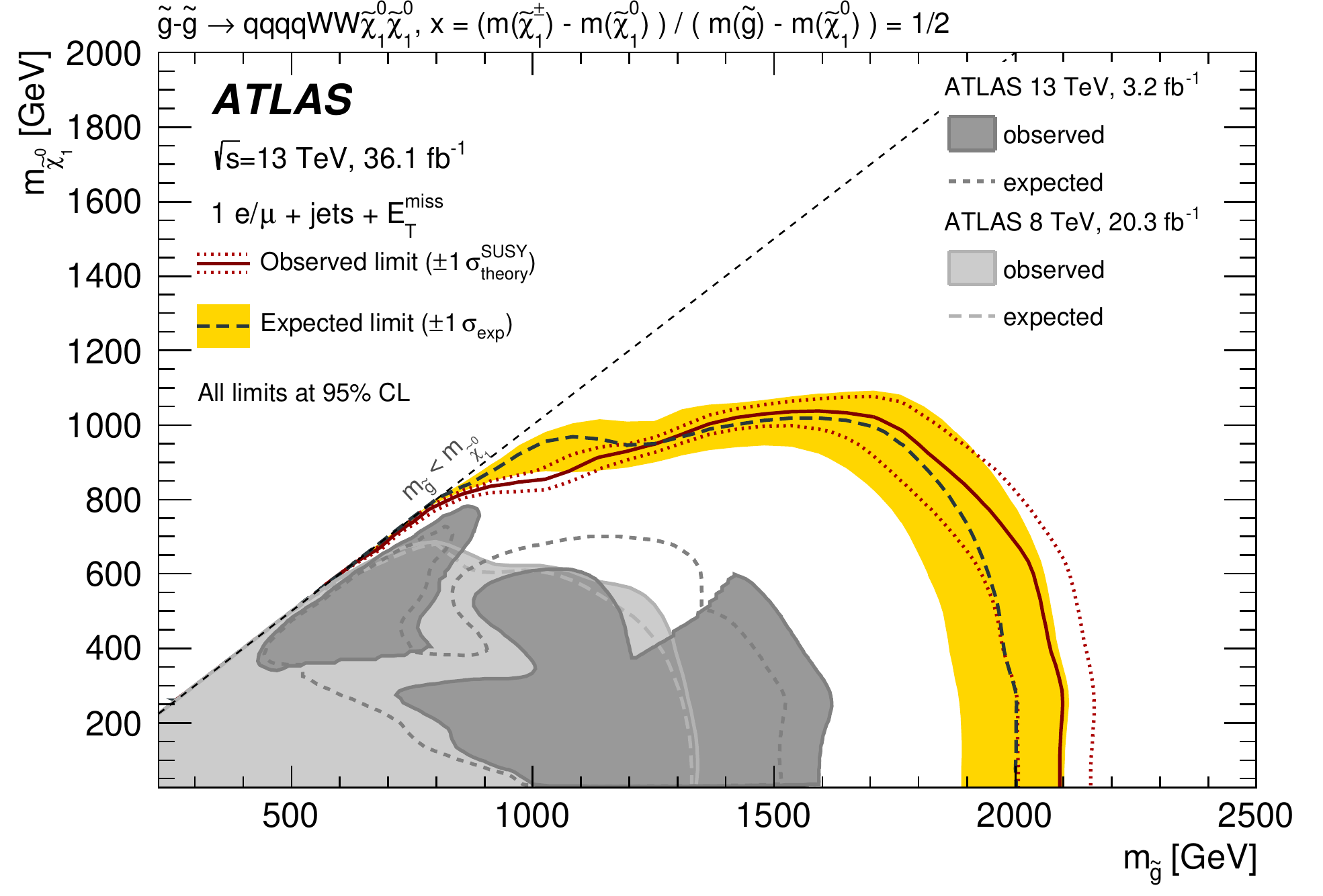}
  \includegraphics[width=0.49\textwidth]{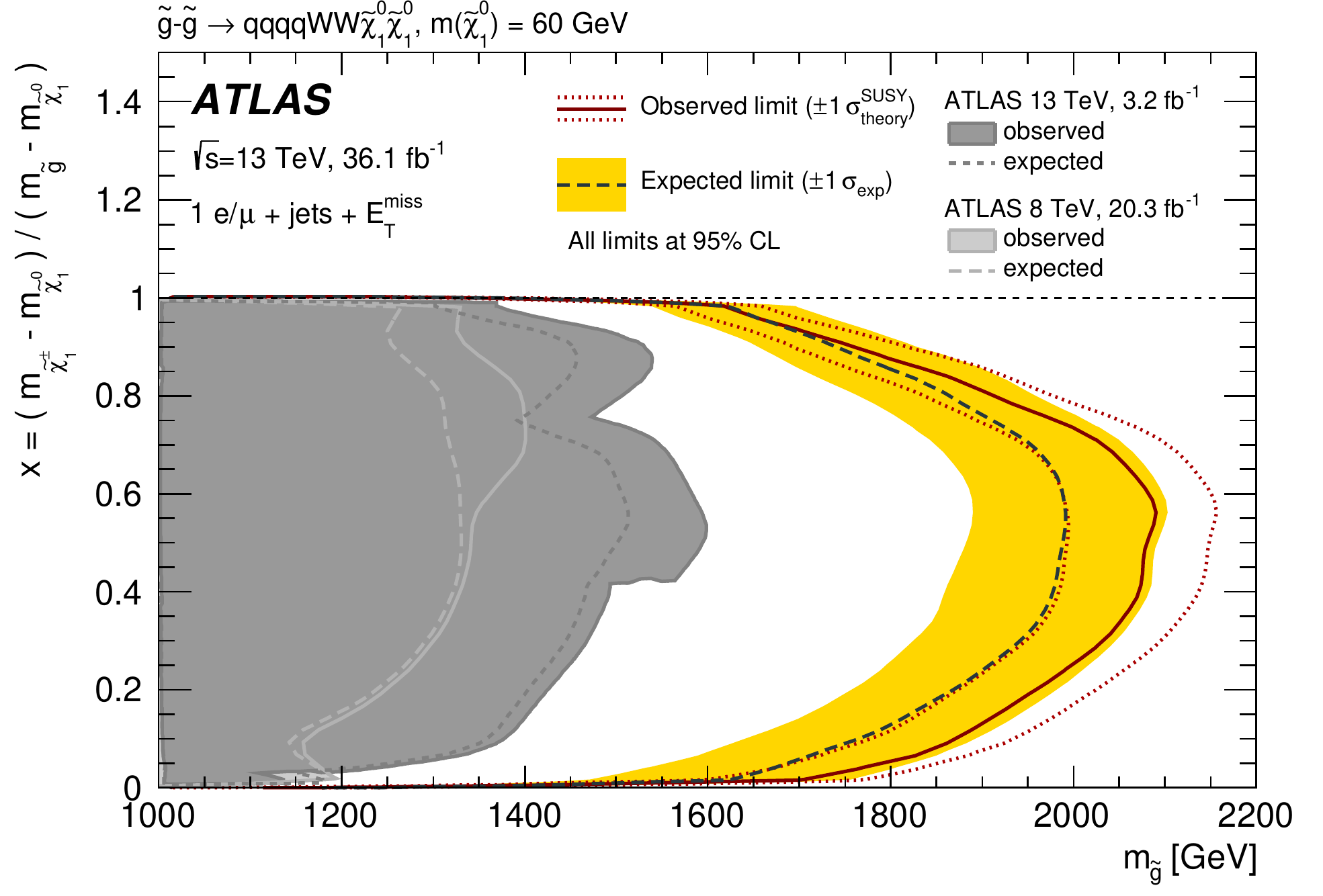}
  \includegraphics[width=0.49\textwidth]{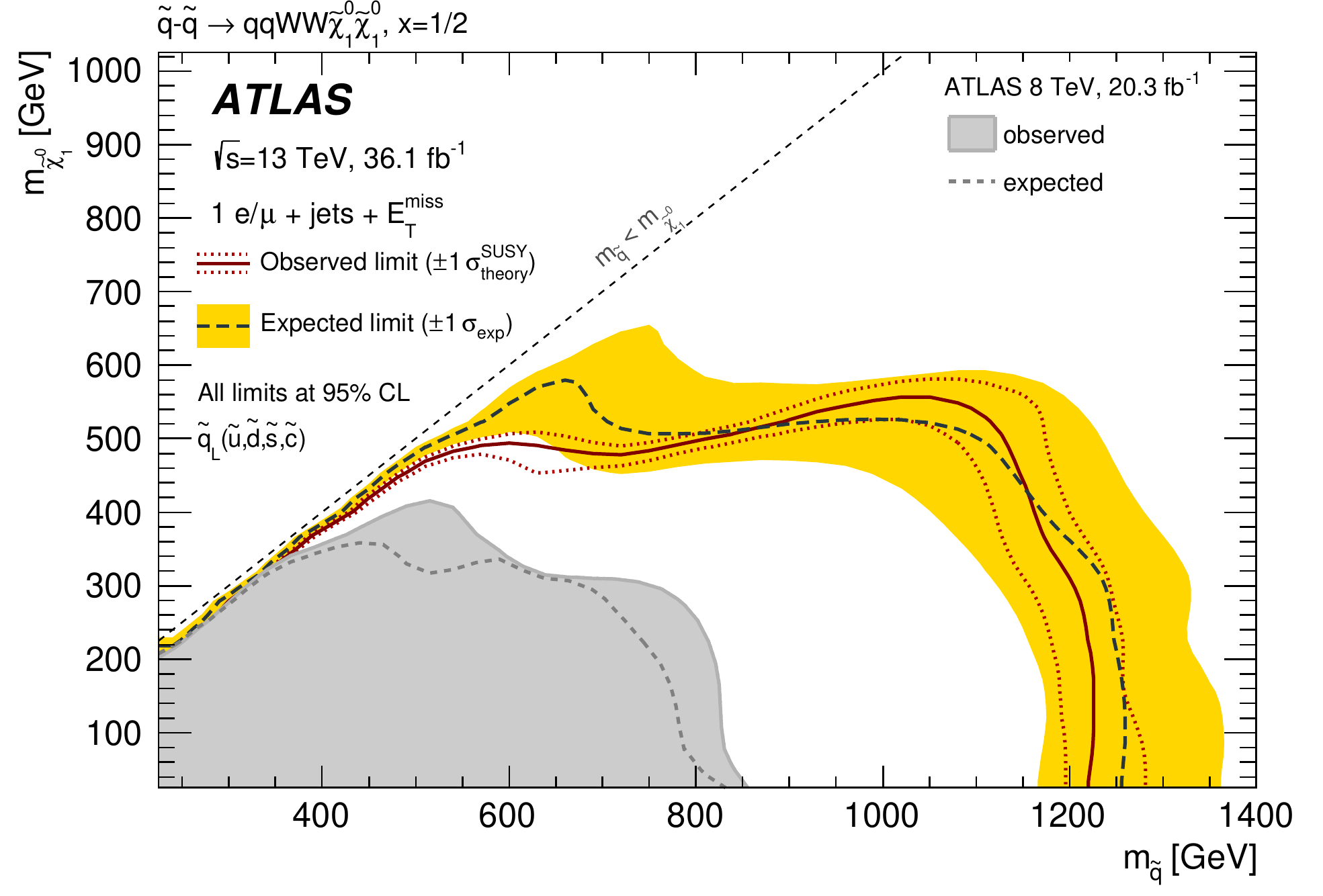}
  \includegraphics[width=0.49\textwidth]{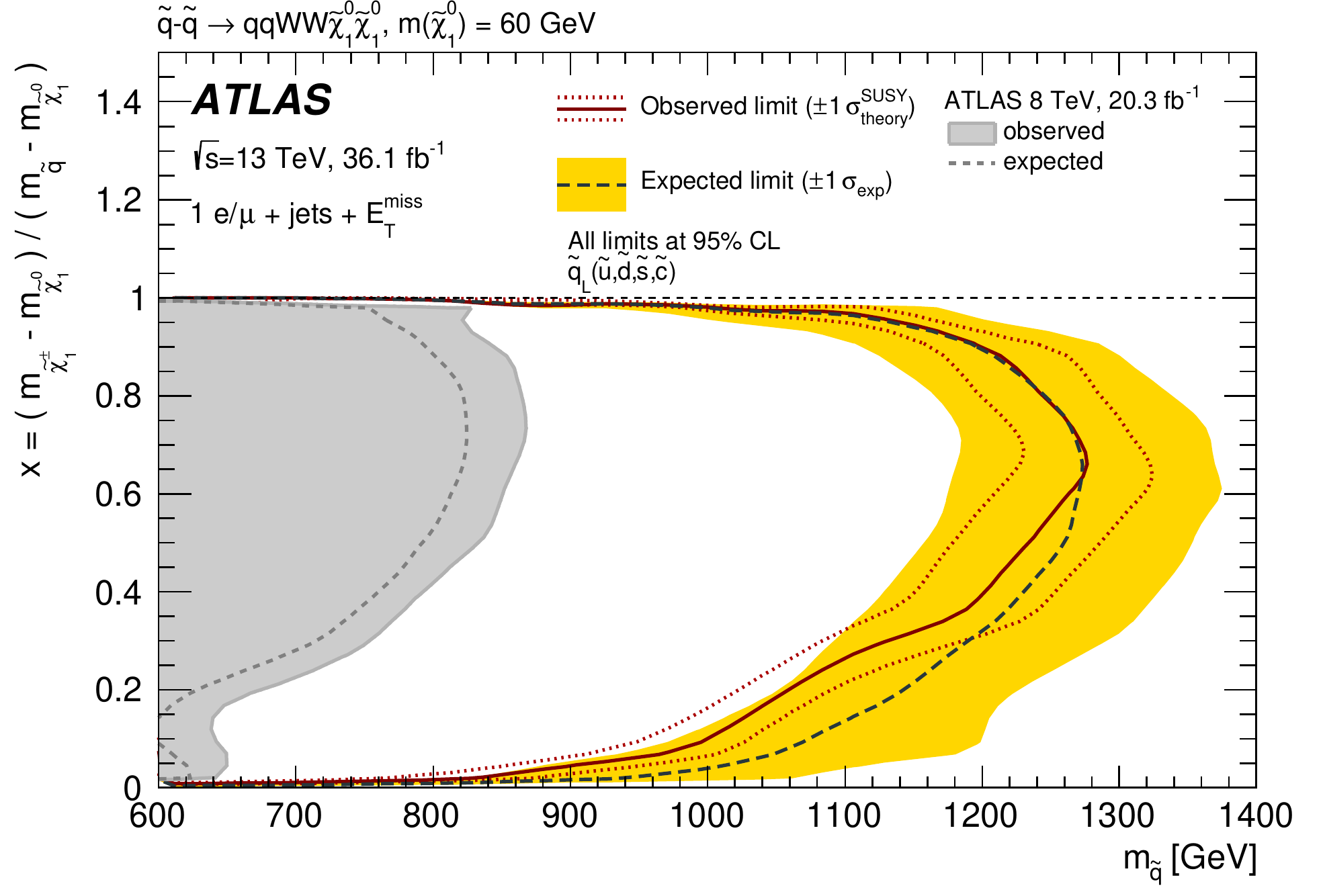}
  \includegraphics[width=0.49\textwidth]{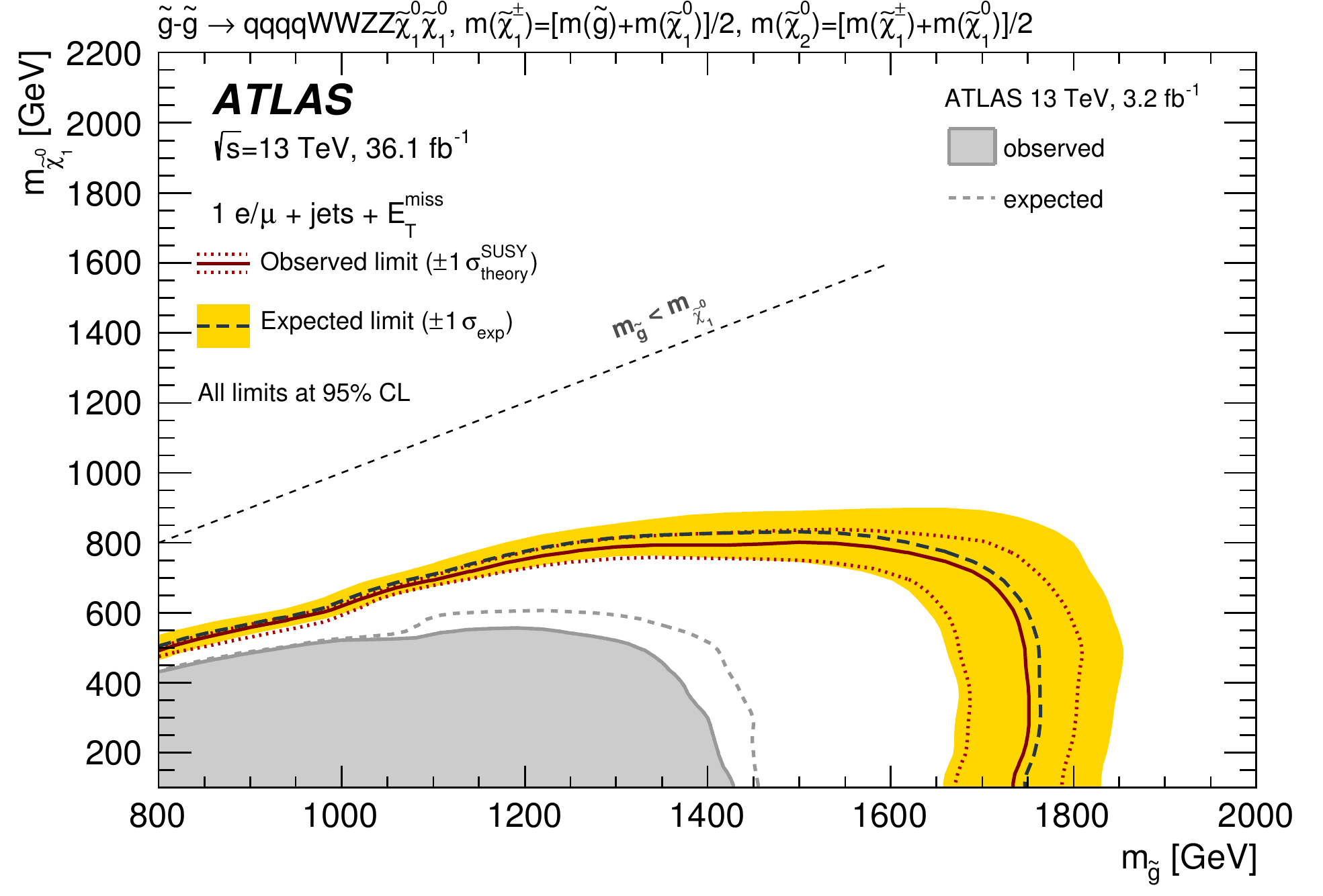}
  \includegraphics[width=0.49\textwidth]{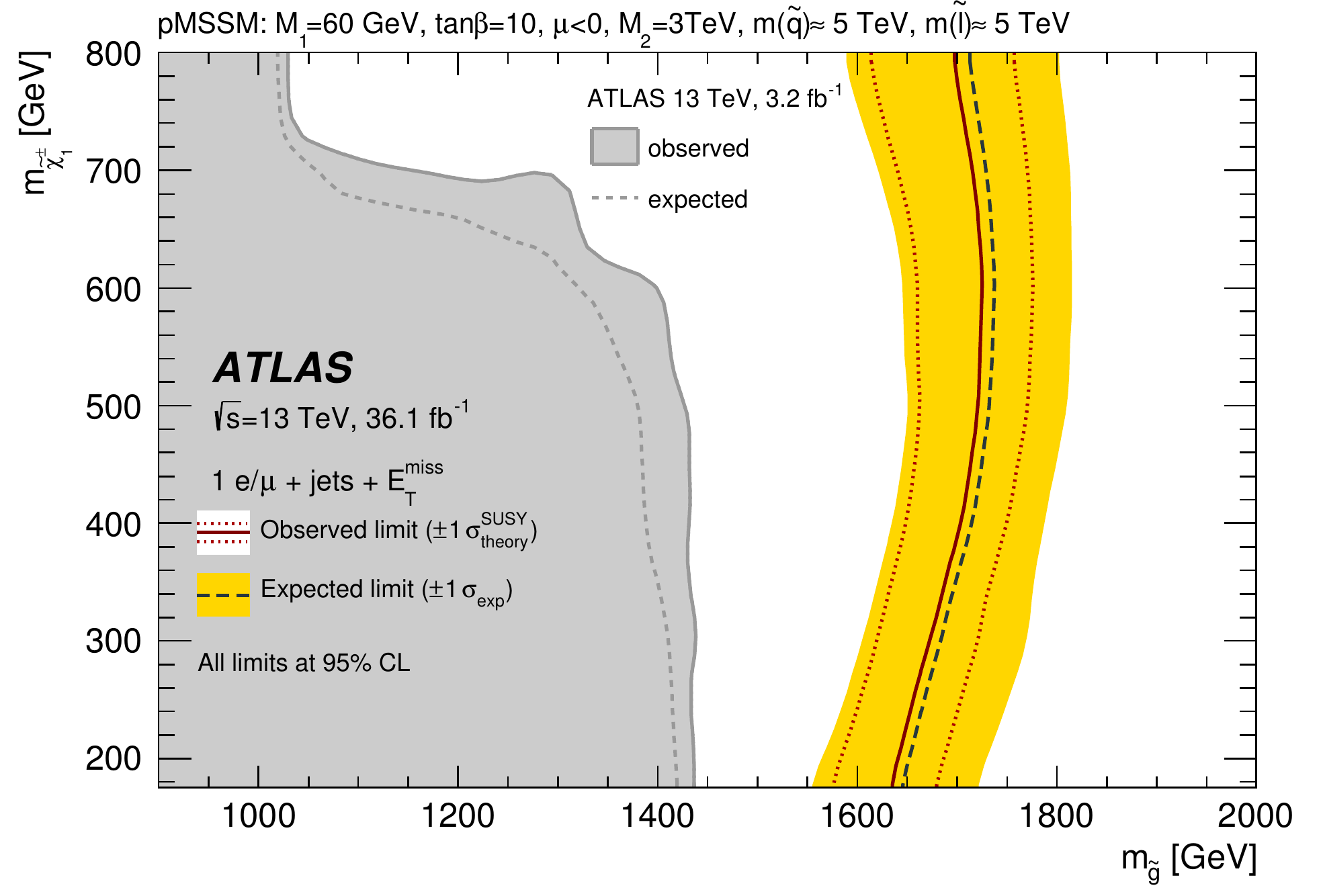}
  \end{center}
\caption{Exclusion
contours for gluino one-step $x=1/2$ (top left), gluino one-step variable-$x$ (top
right), squark one-step $x=1/2$ (middle left) and squark one-step variable-$x$
(middle right), gluino two-step (bottom left), and the pMSSM scenario (bottom right).
The red solid line corresponds to the observed limit with the red dotted lines indicating
 the $\pm 1 \sigma$ variation of this limit due to the effect of theoretical
 scale and PDF uncertainties in the signal cross-section. The dark gray dashed line indicates the
 expected limit with the yellow band representing the $\pm$ 1 $\sigma$ variation of
 the median expected limit due to the experimental and theoretical uncertainties. 
For reference, exclusion bounds from previous searches
with 20.3~\ifb at 8~\TeV~center-of-mass energy~\cite{Aad:2015iea} and 3.2~\ifb
at 13~\TeV~center-of-mass energy~\cite{Aad:2016qqk,Aad:2016jxj} are overlaid
where applicable by the gray area (the observed limit is shown by the solid line, while the dashed line
shows the expected limit) .}
 \label{sec:results:contours} 
\end{figure}

\clearpage

\section{Conclusion}
\label{sec:conclusion}

A search for the pair production of squarks and gluinos in proton--proton
collisions provided by the LHC at a center-of-mass energy of $\sqrt{s}=13\TeV$ 
has been performed by the ATLAS Collaboration.
Events containing one isolated electron or muon, two or more jets, and large
missing transverse momentum are selected in the data collected in 2015 and 2016,
corresponding to an integrated luminosity of \intlumi.
No significant excess over the Standard Model background prediction is
observed. Exclusion limits are set at 95\% CL on generic new phenomena beyond the Standard
Model and on sparticle masses in a number of specific SUSY scenarios. The
exclusion limits extend up to $2.1\TeV$ in gluino mass and $1.25\TeV$ in squark mass thus significantly improving on the sensitivity of previous searches in
this final state.

\section*{Acknowledgments}
We thank CERN for the very successful operation of the LHC, as well as the
support staff from our institutions without whom ATLAS could not be
operated efficiently.

We acknowledge the support of ANPCyT, Argentina; YerPhI, Armenia; ARC, Australia; BMWFW and FWF, Austria; ANAS, Azerbaijan; SSTC, Belarus; CNPq and FAPESP, Brazil; NSERC, NRC and CFI, Canada; CERN; CONICYT, Chile; CAS, MOST and NSFC, China; COLCIENCIAS, Colombia; MSMT CR, MPO CR and VSC CR, Czech Republic; DNRF and DNSRC, Denmark; IN2P3-CNRS, CEA-DSM/IRFU, France; SRNSF, Georgia; BMBF, HGF, and MPG, Germany; GSRT, Greece; RGC, Hong Kong SAR, China; ISF, I-CORE and Benoziyo Center, Israel; INFN, Italy; MEXT and JSPS, Japan; CNRST, Morocco; NWO, Netherlands; RCN, Norway; MNiSW and NCN, Poland; FCT, Portugal; MNE/IFA, Romania; MES of Russia and NRC KI, Russian Federation; JINR; MESTD, Serbia; MSSR, Slovakia; ARRS and MIZ\v{S}, Slovenia; DST/NRF, South Africa; MINECO, Spain; SRC and Wallenberg Foundation, Sweden; SERI, SNSF and Cantons of Bern and Geneva, Switzerland; MOST, Taiwan; TAEK, Turkey; STFC, United Kingdom; DOE and NSF, United States of America. In addition, individual groups and members have received support from BCKDF, the Canada Council, CANARIE, CRC, Compute Canada, FQRNT, and the Ontario Innovation Trust, Canada; EPLANET, ERC, ERDF, FP7, Horizon 2020 and Marie Sk{\l}odowska-Curie Actions, European Union; Investissements d'Avenir Labex and Idex, ANR, R{\'e}gion Auvergne and Fondation Partager le Savoir, France; DFG and AvH Foundation, Germany; Herakleitos, Thales and Aristeia programmes co-financed by EU-ESF and the Greek NSRF; BSF, GIF and Minerva, Israel; BRF, Norway; CERCA Programme Generalitat de Catalunya, Generalitat Valenciana, Spain; the Royal Society and Leverhulme Trust, United Kingdom.

The crucial computing support from all WLCG partners is acknowledged gratefully, in particular from CERN, the ATLAS Tier-1 facilities at TRIUMF (Canada), NDGF (Denmark, Norway, Sweden), CC-IN2P3 (France), KIT/GridKA (Germany), INFN-CNAF (Italy), NL-T1 (Netherlands), PIC (Spain), ASGC (Taiwan), RAL (UK) and BNL (USA), the Tier-2 facilities worldwide and large non-WLCG resource providers. Major contributors of computing resources are listed in Ref.~\cite{ATL-GEN-PUB-2016-002}.

\printbibliography

\clearpage

\newpage 
\begin{flushleft}
{\Large The ATLAS Collaboration}

\bigskip

M.~Aaboud$^\textrm{\scriptsize 137d}$,
G.~Aad$^\textrm{\scriptsize 88}$,
B.~Abbott$^\textrm{\scriptsize 115}$,
O.~Abdinov$^\textrm{\scriptsize 12}$$^{,*}$,
B.~Abeloos$^\textrm{\scriptsize 119}$,
S.H.~Abidi$^\textrm{\scriptsize 161}$,
O.S.~AbouZeid$^\textrm{\scriptsize 139}$,
N.L.~Abraham$^\textrm{\scriptsize 151}$,
H.~Abramowicz$^\textrm{\scriptsize 155}$,
H.~Abreu$^\textrm{\scriptsize 154}$
% ...
\bigskip
\\
% ...
$^{12}$ Institute of Physics, Azerbaijan Academy of Sciences, Baku, Azerbaijan\\
% ...
$^{88}$ CPPM, Aix-Marseille Universit{\'e} and CNRS/IN2P3, Marseille, France\\
% ...
$^{115}$ Homer L. Dodge Department of Physics and Astronomy, University of Oklahoma, Norman OK, United States of America\\
% ...
$^{119}$ LAL, Univ. Paris-Sud, CNRS/IN2P3, Universit{\'e} Paris-Saclay, Orsay, France\\
% ...
$^{137}$ $^{(a)}$ Facult{\'e} des Sciences Ain Chock, R{\'e}seau Universitaire de Physique des Hautes Energies - Universit{\'e} Hassan II, Casablanca; $^{(b)}$ Centre National de l'Energie des Sciences Techniques Nucleaires, Rabat; $^{(c)}$ Facult{\'e} des Sciences Semlalia, Universit{\'e} Cadi Ayyad, LPHEA-Marrakech; $^{(d)}$ Facult{\'e} des Sciences, Universit{\'e} Mohamed Premier and LPTPM, Oujda; $^{(e)}$ Facult{\'e} des sciences, Universit{\'e} Mohammed V, Rabat, Morocco\\
% ...
$^{139}$ Santa Cruz Institute for Particle Physics, University of California Santa Cruz, Santa Cruz CA, United States of America\\
% ...
$^{151}$ Department of Physics and Astronomy, University of Sussex, Brighton, United Kingdom\\
% ...
$^{154}$ Department of Physics, Technion: Israel Institute of Technology, Haifa, Israel\\
$^{155}$ Raymond and Beverly Sackler School of Physics and Astronomy, Tel Aviv University, Tel Aviv, Israel\\
% ...
$^{161}$ Department of Physics, University of Toronto, Toronto ON, Canada\\
% ...
$^{a}$ Also at Department of Physics, King's College London, London, United Kingdom\\
$^{b}$ Also at Institute of Physics, Azerbaijan Academy of Sciences, Baku, Azerbaijan\\
$^{c}$ Also at Novosibirsk State University, Novosibirsk, Russia\\
$^{d}$ Also at TRIUMF, Vancouver BC, Canada\\
$^{e}$ Also at Department of Physics {\&} Astronomy, University of Louisville, Louisville, KY, United States of America\\
% ...
$^{*}$ Deceased
\end{flushleft}


\end{document}